\documentclass[fleqn,usenatbib]{mnras}

\usepackage[T1]{fontenc}
\usepackage{ae,aecompl}

\usepackage{graphicx}	
\usepackage{amsmath}	
\usepackage{amssymb}	

\usepackage{hyperref}
\usepackage{cleveref}
\usepackage{booktabs}
\usepackage{multirow}
\usepackage[dvipsnames]{xcolor}
\usepackage[normalem]{ulem}

\definecolor{darkgreen}{rgb}{1.0, 0.5, 1.0}

\newcommand*{\ksM}{\text{km/s Mpc$^{-1} $}}

\newcommand{\logV}{\log_{10}(V^{\rm{Pivot}})}
\newcommand{\MI}{M_{\rm I}^{\rm{Pivot}}}
\newcommand{\mI}{m_{\rm I}}
\newcommand{\Mi}{M_{\rm I}}
\newcommand{\MD}{M_{\rm D}}
\newcommand{\Ropt}{R_{\rm opt}}
\newcommand{\RD}{R_{\rm D}}
\newcommand{\Hzero}{H_{0}}
\newcommand{\DHzero}{\Delta H_0/H_0}
\newcommand{\DL}{D_{\rm L}}
\newcommand{\Vrec}{V_{\rm rec}}
\newcommand{\Mpc}{{\rm Mpc}}
\newcommand{\kpc}{{\rm kpc}}

\newcommand{\LCDM}{$\Lambda$CDM }

\defcitealias{Riess:2020fzl}{R20}
\defcitealias{Persic:1995ru}{PSS95}
\defcitealias{Yegorova:2006wv}{YS07}

\title[]{Radial Tully-Fisher relation and the local variance of Hubble parameter}

\author[B. S. Haridasu et al]{
Balakrishna S. Haridasu,$^{1,2,3}$\thanks{sharidas@sissa.it}
Paolo Salucci,$^{1,2}$\thanks{salucci@sissa.it}
Gauri Sharma,$^{1,2,3,4,5,6}$\\ 
$^{1}$SISSA-International School for Advanced Studies, Via Bonomea 265, 34136 Trieste, Italy\\
$^{2}$INFN, Sezione di Trieste, Via Valerio 2, I-34127 Trieste, Italy\\
$^{3}$IFPU, Institute for Fundamental Physics of the Universe, via Beirut 2, 34151 Trieste, Italy\\
$^{4}$Department of Physics and Astronomy, University of the Western Cape, Cape Town 7535, South Africa \\
$^{5}$University of Strasbourg, CNRS UMR 7550, Observatoire astronomique de Strasbourg, F-67000 Strasbourg, France \\
$^{6}$Department of Physics and Astronomy, University of the Western Cape, Cape Town 7535, South Africa
}

\date{Accepted XXX. Received YYY; in original form ZZZ}

\pubyear{2021}

\begin{document}
\label{firstpage}
\pagerange{\pageref{firstpage}--\pageref{lastpage}}
\maketitle


\begin{abstract}

Utilizing the well-established Radial Tully-Fisher (RTF) relation observed in a `large' (843) sample of local galaxies, we report the maximum allowed variance in the Hubble parameter, $\Hzero$. We estimate the total intrinsic scatter in the magnitude of the RTF relation(s) implementing a cosmological model-independent cosmographic expansion. We find that the maximum allowed local variation in our baseline analysis, using 4 RTF relations in the galaxy sample is $\DHzero \lesssim 3 \%$ at a $95\%$ C.L. significance. Which is implied form a constraint of $\DHzero = 0.54^{+1.32}_{-1.37} \%$ estimated at $\DL\sim 10\, [\Mpc]$. Using only one `best-constrained' radial bin we report a conservative $95\%$ C.L. limit of $\DHzero \lesssim 4 \%$. Through our estimate of maximum variation, we propose a novel method to validate several late-time/local modifications put forth to alleviate the $H_0$ tension. We find that within the range of the current galaxy sample redshift distribution $10 \, [\Mpc] \le \DL \le 140\, [\Mpc]$, it is highly unlikely to obtain a variation of $\DHzero \sim 9\%$, necessary to alleviate the $H_0$-tension. However, we also elaborate on the possible alternative inferences when the innermost radial bin is included in the analysis. Alongside the primary analysis of fitting the individual RTF relations independently, we propose and perform a joint analysis of the RTF relations useful to create a pseudo-standardizable sample of galaxies. We also test for the spatial variation of $H_0$, finding that the current samples' galaxies distributed only in the southern hemisphere support the null hypothesis of isotropy, within the allowed noise levels. 

\end{abstract}




\section{Introduction}
\label{sec:Introduction}

To determine the value of the ``Hubble Parameter": $H(z) $ at different redshifts and in particular at present ($\Hzero$) has become one of the most important and telling cosmological measurements. The well-established and increasingly prominent $\Hzero$-tension   \citep{Verde:2019ivm, DiValentino:2020zio} has paved the way to speculate several modifications to the concordance model of cosmology. At present, the significance of this discrepancy between the local model-independent Cepheid calibration-based supernovae yield $\Hzero = 73.04 \pm 1.04\, \ksM$ \citep{Riess:2020fzl} and the Cosmic Microwave Background (CMB) based model-dependent (\LCDM) indirect estimate $\Hzero = 67.66\pm 0.49\, \ksM$ from the improved PR4 analysis in \cite{Tristram:2023haj} is about $\sim 5\sigma$. In addition, the latter CMB estimate is corroborated by the Baryon Acoustic Oscillation data \citep{Alam_2021, Bourboux20, Zhao18_dr14}, providing $H_0 = 67.81\pm0.38\, \ksM$ which only increases the significance of the said tension. A more recent claim for a $8.2\sigma$ tension was presented in \cite{Riess:2024ohe}, addressing the cepheid crowding and their power-luminosity relation using JWST observations. Several reviews and discussions now provide a very good overview of the state of the tension \citep{Abdalla:2022yfr, Riess:2019qba, Perivolaropoulos:2021jda, Schoneberg:2021qvd, Shah:2021onj,  Hu:2023jqc, Efstathiou:2020wxn, Freedman:2021ahq, Knox20, Verde:2023lmm, Akarsu:2024qiq}. Also, other local calibration techniques and measurements either provide a similar disagreement with the CMB estimate or at least do not yield immediate resolutions to explore \citep{Pesce:2020xfe, deJaeger:2020zpb, Schombert:2020pxm, Chen:2019ejq, Bonvin:2016crt, Yuan:2019npk, Blakeslee:2021rqi, Freedman:2019jwv, Wong:2019kwg, Jee:2019hah, DES:2019fny, Huang:2019yhh}.

To resolve longstanding Hubble tension, several approaches have been proposed and explored ranging from modifications to early-universe physics \citep{Poulin:2021bjr, Karwal16, Niedermann:2019olb, Niedermann:2020qbw, Zhao:2017urm, delaMacorra:2021hoh, Jedamzik:2020zmd, Kreisch:2019yzn} to late-time/intermediate redshift physics \citep{Nygaard:2023gel, Vattis:2019efj, Akarsu:2019hmw, Blinov:2020uvz, Anchordoqui:2020djl, Haridasu:2020xaa,  Tutusaus19, Sola:2017znb, Tutusaus:2023cms, Khosravi17, Adil:2023exv, Akarsu:2023mfb, Lapi:2023plb}, local Universe \citep{Keenan13, Whitbourn14, Shanks19, Hoscheit17, Colgain18, Lukovic19, Cai:2020tpy, Alestas:2020zol, Kenworthy_2019}, leading to extended discussions \citep{Lee:2022gzh, Addison:2017fdm, Poulin:2020xcz, Haridasu:2020pms, Vagnozzi:2023nrq, Bernal:2021yli, Krishnan:2020vaf, Vagnozzi:2019ezj, Gomez-Valent:2023uof, Cao:2023eja, Gomez-Valent:2023uof}. Alongside modifications to the physics of the Universe, possibilities that tension can arise due to systematics have been explored many times \citep[e.g.,][]{Mortsell:2021tcx, Mortsell:2021nzg}. Several techniques to study the possible resolutions of the Hubble tension under the least possible cosmological assumptions have also driven the need for model-independent techniques that have substantial significance in recent times, for example, \cite[refernces therin]{ Liao:2020zko, Pandey:2019yic, Haridasu18_GP, Gomez-Valent18, Lemos18, Lyu20, Liu:2022bmn, Qi:2023oxv, Du:2023zsz, Li:2024elb}. 

In this context, the Tully-Fisher \citep{Kourkchi:2020iyz} and the Baryionic Tully-Fisher \citep{2022MNRAS.511.6160K} relations, that have the advantage of possessing a clear physical justification (see \cite{1993MNRAS.262..392S, Salucci:2018hqu}), have been utilized to estimate the local expansion rate, which has been consistent and at times providing even larger values of $H_0$ with respect to the SH0ES estimate using SNe. These methods essentially highlight the necessity of alternate local estimations of $\Hzero$ aiding immensely the discussion on $\Hzero$-tension and providing robust support to the Chepehid-SNe-based local determination of the same. For a recent discussion see also \cite{Tully:2023bmr}. Along these lines, we intend to introduce the possibility of assessing the same using the Radial Tully-Fisher (\textbf{RTF}) relation for the first time. The RTF relation has been well established in \citet{Yegorova:2006wv}, following the discovery of a strong global relationship between the rotation velocities and the absolute magnitudes ($\Mi$) of the $\sim 800$ nearby galaxies \citep[and references therein]{Persic:1995ru, Salucci:2018hqu}. The RTF relation indicates that there exist independent radial Tully-Fisher-like relations at different galactocentric radii within spiral galaxies. In the current work, we take advantage of the RTF relation to obtain limits in the allowed variation of the Hubble rate within the local Universe $z< 0.035$, more precisely at the edge of the `Hubble-Flow'. 

Several local ($z< 0.1$) and ultra-local ($z< 0.01$) (\cite{Alestas:2020zol, Marra:2021fvf, Desmond:2020wep, Desmond:2019ygn, Benevento:2020fev, Banerjee:2023opo}) physical resolutions and possible variations in the standardization of SNe (\cite{Alestas:2021luu, Benisty:2023laj, Camarena:2023rsd, Ruchika:2023ugh, Perivolaropoulos:2023iqj}), have also been suggested to alleviate the $\Hzero$-tension, essentially relying on the fast transitions of physics in addition to the local void \cite{Keenan13, Hoscheit17, Whitbourn14} and extremely local ($z< 0.015$) sharp transition of the dark energy equation of state \cite{Camarena19} etc. We anticipate the ability of the RTF relation to constrain proposed local modifications to the cosmological scenario by estimating the allowed variance in the local estimation of the Hubble parameter. We begin by reanalyzing the RTF relations presented in \cite{Persic:1995ru} (here onwards \citetalias{Persic:1995ru}) and \cite{ Yegorova:2006wv} (here onwards \citetalias{Yegorova:2006wv}) using improved Bayesian techniques. Given the nature of the empirical relations, we also propose a methodology improving the RTF relations by introducing a joint analysis to constrain the relations simultaneously, modeling a covariance among the individual relations. 

The organization of the paper is as follows. After a brief introduction to the Radial Tully-Fisher relation in \Cref{sec:RTF}, in \Cref{sec:Data} we present the data. The cosmographic methodology and fitting of the RTF relations is described in \Cref{sec:method}. Finally, in \Cref{sec:Results}, we present the results with an extended discussion.

\section{Radial Tully-Fisher relation}
\label{sec:RTF}

Adopting $\Ropt$, the radius encompassing 83\% of the total light of a Spiral Galaxy, as the reference size of its stellar disk, the radial Tully-Fisher relation is a family of TF-like relations observed in disk systems between the galaxy absolute Magnitude in a certain $j$ band (e.g. $\Mi$, the infrared band) and the rotational velocity $V(R/\Ropt)$ measured at fixed normalized radii $R/\Ropt$. These relations have been well established in \citetalias{Yegorova:2006wv} with the help of the large sample of galaxies with good quality RCs from \citetalias{Persic:1995ru} and that of two additional samples with 86 and 81 high-quality RCs \cite{Courteau1997,Vogt2004}. In \cite{Fontaine:2018sby} the RTF relationships have been established for a sample of 36 Dwarf Spirals and Irregulars. 

The RTF relations, for a given magnitude $M_j$, are represented by a class of linear models,
\begin{equation}
 \label{eqn:RTF}
     M_j = a_n\times \log_{10}(V_{n}) + b_n
\end{equation}
where the subscript $n$ tags the radial bins in which the RTF is established. These bins are centered at the radii $R_n$, defined in terms of fractions of the optical radii. For the \citetalias{Persic:1995ru} sample, that we use in this work: $R_n \equiv n/5 \Ropt$ and the bin width is $1/5 \Ropt$. Noticeably, the  RTF has emerged also by adopting a smaller bin size, e.g. $1/15 \Ropt $, as in \citetalias{Yegorova:2006wv} for the \cite{Courteau1997,Vogt2004} samples of high-resolution RCs. 

For the galaxies of the current Sample,  $V_n \equiv V(R_n)$ is the average value of the velocity data in each $n^{\rm th}$ bin, $a_n$ and $b_n$ are the slope and the intercept of the linear RTF relations found for the data belonging to the $n^{\rm{th}}$ bin\footnote{Since the surface luminosity density of the stellar disk $I(R)$ is very similar in all spirals and it is represented by the well-known Freeman profile:  $I(R) \propto e^{-R/\RD}$, the lengthscale $\Ropt = 3.2 \RD$ describes the distribution of luminous matter for every spiral galaxy in the same consistent way.}. Let us notice that in the constant $b_n$ we can  
incorporate the term $ M_j(\infty)-M_j(R_n)$ which at a fixed $n$ is equal in all galaxies, so that the L.H.S of \Cref{eqn:RTF} can be interpreted as the aperture $j$-magnitude at the radius $R_n$, a quantity related to the mass in stars inside such radius. 

\begin{figure}
    \centering
\includegraphics[scale=0.59]{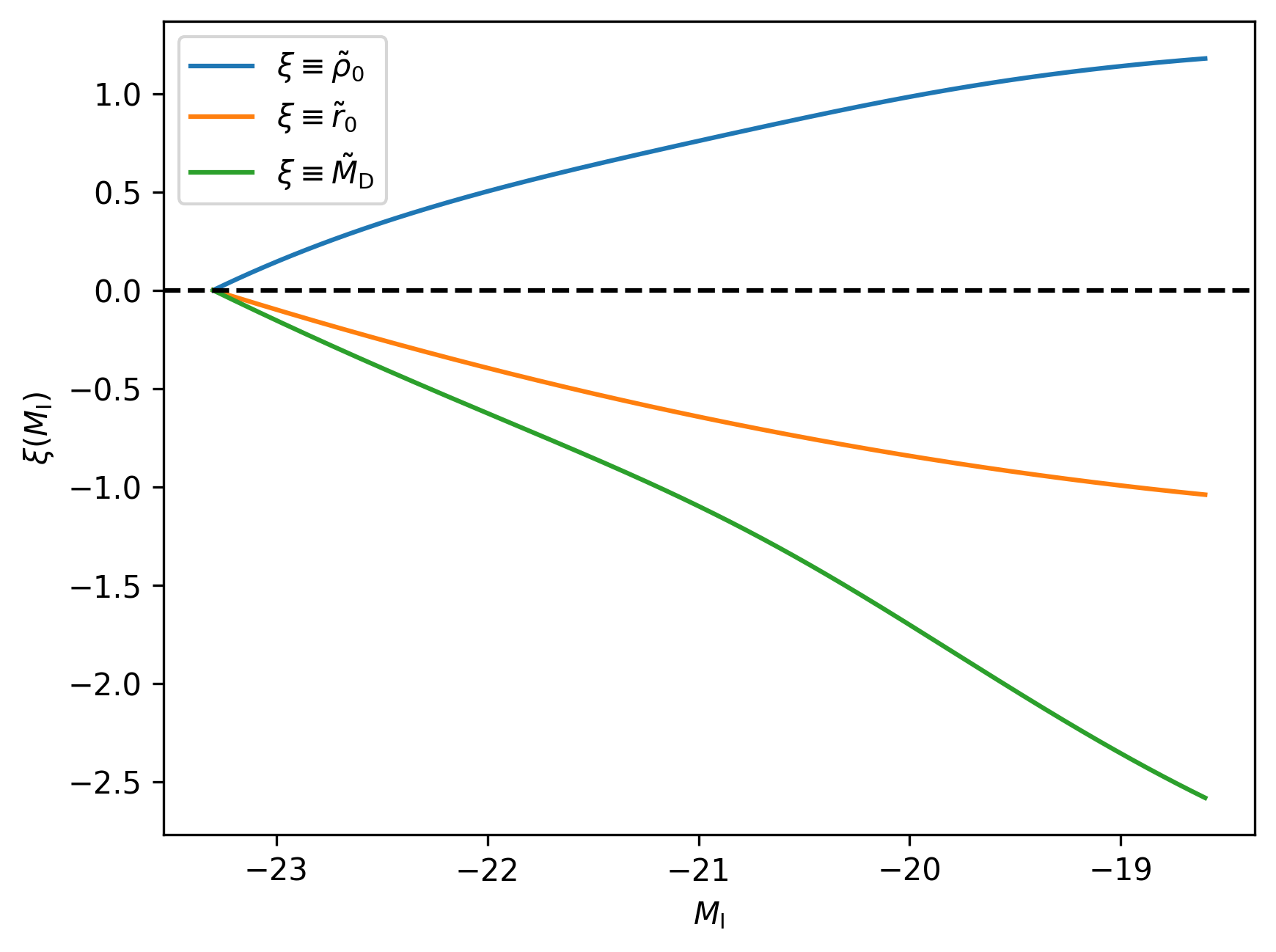}
\caption{The parameters of the mass model of Spiral galaxies plotted as functions of the infrared Magnitude $\Mi$. $\tilde\rho_0(\Mi)= \rho_0(\Mi)/(1.2\times 10^{-25} \rm{g/cm^3} )$ (blue), $\tilde r_0(\Mi)=r_0(\Mi)/(54 \, \kpc)$ (orange), $\tilde \MD(\Mi)= \MD(\Mi)/(4 \times 10^{11} M_\odot)$ (green).
}
\label{fig:TF_parameters}
\end{figure}

It is worth demonstrating here that the RTF relationship with the features described above has a strong physical background as the originating TF one: it is a direct consequence of the fact that spiral galaxies are rotationally supported; their circular velocities, at a radius $R$ balance the gravitational attraction of the galaxy mass inside this radius $R$. Remarkably, in Spirals the mass distribution has universal features (see \cite{Salucci:2018hqu}) and it includes 1) a Freeman stellar disk of mass $\MD$ and length-scale $\RD(\Mi)$, whose contribution to the circular velocity $V(R)$ can be written as $V_{\rm d}(R/\RD;\MD)$ with the disk mass $\MD$ as a free parameter of the mass model. 2) a cored dark matter halo whose contribution to $V(R)$ can be written as  $V_{\rm h}(R;\rho_0,r_0) $ (\cite{Salucci:2007et}), with the central density $\rho_0$ and the  core radius $r_0$ also as free parameters of the mass model. The resulting circular velocity model:  
\begin{equation}
\label{eqn:Vmodel}
V_{\rm model}(R)=\left[V^2_{\rm d}(R/\RD;\MD)+V^2_{\rm h}(R;\rho_0,r_0)\right]^{1/2}
\end{equation}
successfully fits the circular velocities $V(R)$ of the entire family of Spiral galaxies  (\citetalias{Persic:1995ru}; \cite{Salucci:2000ps, 2017MNRAS.465.4703K, Salucci:2018hqu}) provided that the above three free structural parameters become specific functions of the galaxy's infrared magnitude $\Mi$ 
\begin{equation}
\label{eqn:struct_params}
M_D(M_I); \ \ \ r_0(M_I); \ \ \ \rho_0(M_I),
\end{equation}
shown in \Cref{fig:TF_parameters} and given in \citetalias{Persic:1995ru} (see also \cite{Salucci:2000ps}). By inserting the \Cref{eqn:struct_params} into \Cref{eqn:Vmodel}, one obtains that the latter becomes equivalent to the \Cref{eqn:RTF} with the values of parameters given by \Cref{tab:LRS} (see \citetalias{Yegorova:2006wv} for the more details).

A further prediction of the physical origin of the RTF relationship is that in the innermost bin (i.e. for  $n=1$) the relation should show a scatter sensibly larger than in the outer bins (i.e. for  $2\leq n \leq 5$). This is due to the presence, in a good fraction of the objects of the sample, of a central stellar bulge that provides an important contribution to the mass enclosed in the innermost radius. The mass of this spheroid has a trend with that of the stellar disk, so that, the RTF continues to exist also for $n=1$, but with a scatter larger than those at farther radii, not affected by the central bulge mass (see \citetalias{Yegorova:2006wv}). Let us also point out that, according to the above  velocity model and directly supported by RCs data, we have
\begin{equation} 
V_{\rm model} (n/5\,\Ropt,-23.) \simeq 10^{2.5} \ \rm{ km/s }
\end{equation}
for  $1\leq n \leq 5$. Thus the RTF relationship, that will be used here as a distance indicator, reflects the condition of the centrifugal equilibrium of a stellar disk embedded in a dark halo.

\label{sec:Data}

\begin{figure}
    \centering
    \includegraphics[scale=0.58]{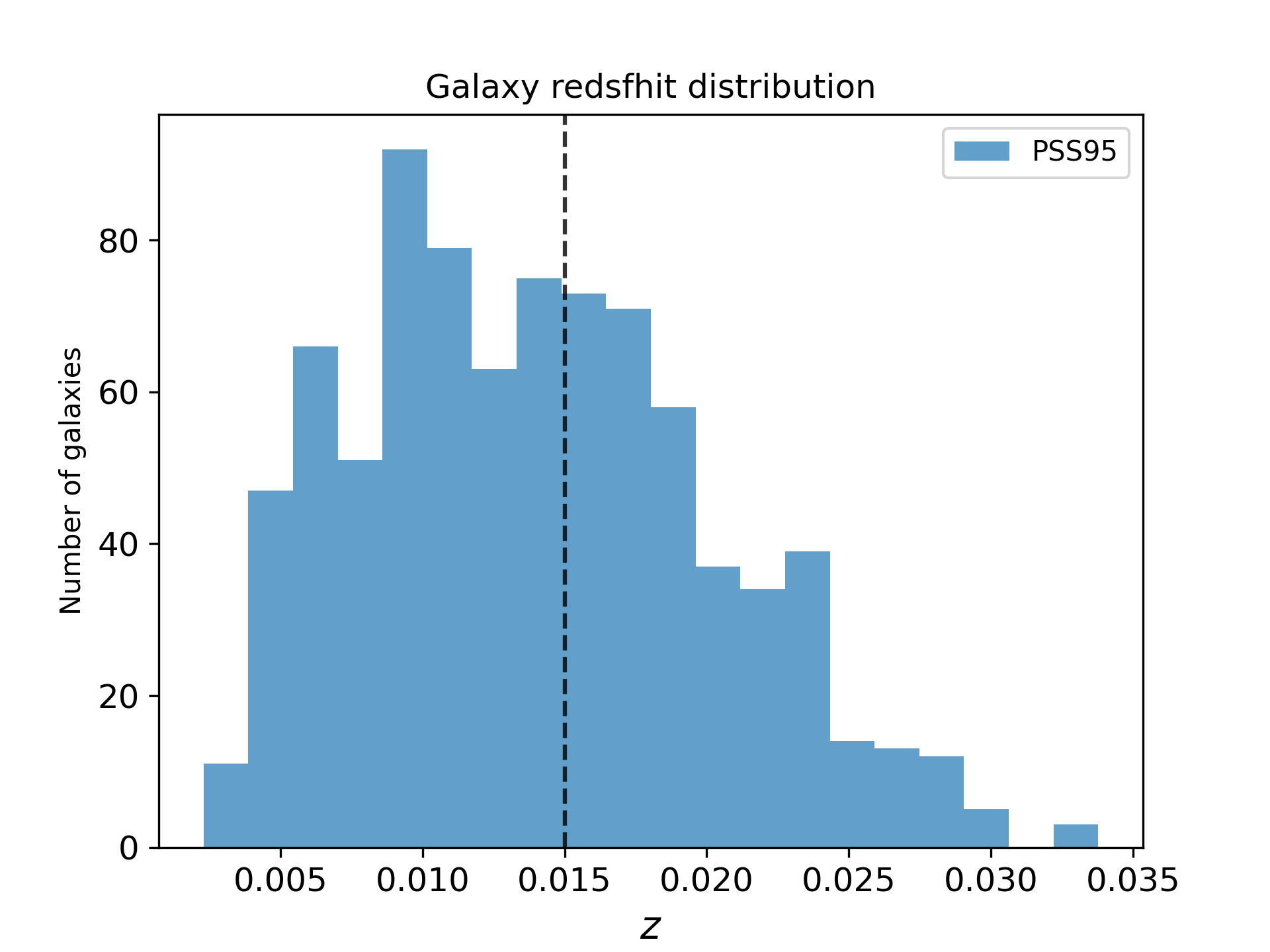}
    \caption{Redshift distribution of the \citetalias{Persic:1995ru} sample consisting of 489 galaxies below $z<0.015$ and 384 galaxies from $z>0.015$.  }
    \label{fig:TF_redshift_dist}
\end{figure}

\textit{Data:} In the current work, we utilize the same dataset that has been used to work out the `Universal rotational curve' in \citetalias{Persic:1995ru} and later analyzed in \cite{Yegorova:2006wv} to work out the Radial Tully Fisher. The dataset provides the magnitude of the galaxies in the `I' band against the binned rotational velocity in each of the optical bins. In \Cref{fig:TF_redshift_dist} we show the redshift distribution of the galaxies which range between $0.005 \leq z \leq 0.035$, incidentally centered around $z\sim 0.015$, similar to the lower limit of $z>0.023$ usually taken to estimate the local value of $\Hzero$ in SH0ES analysis \cite{Riess:2016jrr, Riess_2018}, allowing the SNe to be in the `Hubble flow'. 
Notice that the peculiar motions of the galaxies in our sample, assumed to be $200\, \rm{km/s}$, affect the determination of their redshift with an error, on average, of about $\pm 0.08$ mag, i.e. a value smaller than the intrinsic scatter of the RTF relations, that amount to  $0.12-0.25$ mag; therefore, systematics that may arise in having not detailed these motions is small and likely washed out by the random uncertainty of the RTF relations. Moreover, in this work, the  RTF distance indicator is used for tasks, whose complement does not require a good knowledge of the peculiar motions of galaxies.      

In \Cref{fig:PS_95} we show the RTFs for the different optical radii, alongside the best-fit linear relations, elaborated in \Cref{sec:Results}. The distribution of these galaxies in the sky is shown in the \Cref{fig:sky}, which are a subset of those presented in the southern sky survey \cite{Mathewson:1992ia}. Here we show the distribution of the galaxies in \Cref{fig:sky}, in the J2000 system in contrast to the B1950 as was originally presented in \cite{Mathewson:1992ia}. 

In summary, the Radial TF has been established in the process of investigating, in an innovative way (\citetalias{Yegorova:2006wv}), the mass distribution of Spirals: for the region inside their optical radii essentially providing evidence for: i) the presence of a massive Dark Component ii) the decrease of the DM/total matter fraction with increasing galaxy luminosity iii) a very shallow halo density profile iv) the presence of a central bulge component. \textit{The role of the RTF thus far was therefore primarily intended to study dark matter properties in galaxies (e.g. \cite{Salucci:2018hqu}). However, given the recent Hubble Tension, this tight and physically motivated Tully-Fisher-like relationship, even using the `original' dataset can be a very efficient distance-indicator. The improvement, with respect to the TF is obvious, that the latter uses one circular velocity at a reference radius per galaxy, while the RTF exploits the full RC rotation curve inside the same optical radius of each galaxy.}

\section{Method}
\label{sec:method}

We adopt the linear models shown in \Cref{eqn:RTF} to perform regression to obtain the posteriors for each of the radial bins defined above in terms of the optical radius. In detail, firstly we perform a simple linear regression for each of the sub-samples split into the 6 bins with centers located at each of the first six $0.2$ multiples of $\Ropt$, wherein all the data points are assumed to be independent and no correlation is assumed in the analysis. For consistency, after performing the initial regression we exclude the data points that are more than $3 \sigma$ confidence level away from the posterior regressed line, to exclude the outliers. However, we also note that this exclusion of the outliers does not significantly affect the final inferences drawn from the analysis. The analysis requires 3 parameters for each of the RTF relations taking into account the slope, intercept, and intrinsic scatter $\{a_n, b_n, \sigma_n\}$, amounting to a total of 18 total parameters to fit the 6 optical bins considered in the current dataset. Note that throughout the analysis we assume and fix a fiducial value of $H_0 = 70\, \ksM$ and $q_0 = -0.55$. Varying the value of $H_0$ within the current analysis with a simple cosmographic background, presented in the next section, only amounts to an overall shift of all RTF relations with no implication for the shape of the RTF relations. This remains a valid assumption given no significant empirical indications to go beyond the linear relations. 

\subsection{Cosmography}
\label{sec:CGY}

To utilize the given galaxy samples to estimate the maximum allowed variance in the Hubble parameter we implement a simple cosmographic approach to model the luminosity distance as,

\begin{equation}
\label{eqn:Cosmography}
    \DL(z) = \frac{c}{\Hzero}\times \left[ z + \frac{1}{2}(1-q_0)z^{2}\right],
\end{equation}
where $\Hzero$ is the current expansion rate and $q_0$ is the deceleration parameter. The distance modulus can now be written as the difference between the apparent magnitude and the absolute magnitude,

\begin{equation}
\label{eqn:cosmoMI}
\mI - \Mi = 25 + 5 \log_{10}\left( \DL(z) \left[ {\rm Mpc} \right]\right),
\end{equation}
wherein we utilise the recession velocity ($\Vrec$) of each of the galaxies to obtain the redshift $z = \Vrec/c $. Note that the above equation can further be approximated, taking only the first-order term in \Cref{eqn:Cosmography} into account, 

\begin{equation}
\label{eqn:Cosmography_final}
\mI - \Mi = 25 + 5 \log_{10}\left( \Vrec/ \Hzero\right).
\end{equation}

Finally, given the assumed cosmography to obtain the absolute magnitude of the data points and the posteriors of linear regression models obtained through the fitting relation \Cref{eqn:RTF} we construct the residuals of the absolute magnitude as a function of the redshift. The redshift of the galaxies is consistently obtained utilizing the recession velocity of the galaxies, as described earlier. Also, one could equivalently present the same in terms of the luminosity distance of the galaxies instead of redshift, which we present as the final result (elaborated in \Cref{sec:Results}). 

\subsection{Joint analysis}
\label{sec:JA}
The methodology described so far follows \citetalias{Yegorova:2006wv} and constrains each of the RTF relations individually. While we utilize the same as a first step, in this work we also introduce a method to perform joint analysis and simultaneously constrain the RTF relations. As can be seen in the \Cref{fig:PS_95} and also in Figure 2. of \citetalias{Yegorova:2006wv}, all the individual RTF relations at each optical bin converge at a `pivot' that remains to be a fixed point depending only on the value of $H_0$ assumed in the conversion of the observed apparent magnitudes to the absolute magnitude $\Mi$ through \Cref{eqn:Cosmography}, and the distance modulus expression. As shown in \Cref{sec:RTF} we remind that the mass model of spiral galaxies at all radii is observationally unrelated to the RTF, implying the existence of this pivot quantity. 

This, in turn, modifies the individual RFT description in \Cref{eqn:RTF} with a pivot as,
\begin{equation}
    \label{eqn:RTF_joint}
    M_j - \MI  = a_n\times (\log_{10}(V_{n}) - \logV)
\end{equation}
where the pivot is given by $\{\MI, \logV \}$ and the corresponding slopes $a_n$ of the individual relations. In contrast to the total of 18 parameters describing the 6 independent relations, within the joint analysis setup, we have 14 parameters: 12 describing the slopes and intrinsic scatter of the six relations, and two fixing the pivot. 

Finally, the likelihood of the individual linear regression analysis within each of the radial bins $n$ can be written as, 

\begin{equation}
    -2 \log(\mathcal{L}) = \sum^{N_{\rm gal}^{n}}\left[ \frac{(\Mi^{\rm obs} - \Mi^{\rm theo}(V_{n}))^2}{\sigma_{n}^{2}}\right],
\end{equation}
where the $\Mi^{\rm obs}$ is constructed using \Cref{eqn:cosmoMI}, while the $\Mi^{\rm theo}(V_{n})$ assume the form in \Cref{eqn:RTF}, with the free parameters $\Theta \equiv \{a_n, b_n\}$, and $\sigma_n$ is a free parameter assessing the intrinsic scatter of the data. Here $N_{\rm gal}^{n}$ is the number of galaxies with rotational curve velocities in $n^{\rm th}$ radial bin. Similarly, the likelihood for the joint analysis is written as,

\begin{equation}
    -2 \log{(\mathcal{L})} = \sum^{n} \sum^{N_{\rm gal}^{n}} \left[ \frac{(\Mi^{\rm obs} - \Mi^{\rm theo}(V_{n}))^2}{\sigma_{n}^{2}}\right], 
\end{equation}
wherein a summation over the radial bins $n$ is included and the $\Mi^{\rm theo}(V_{n})$ is given by \Cref{eqn:RTF_joint}, with free parameters $\Theta \equiv \{\MI, \logV, \mathbf{a_n}\}$ amounting to a total of 14 parameters as described earlier.
Note that one could include an additional parameter $\tau$ as a penalty term $\log(\tau + \sigma_n^2)$ to the total likelihood. This would describe a degeneracy/covariance between the 6 RTF relations, to be more conservative and validate the utility of the magnitude used in RTF relation at each optical bin. In other words, the parameter $\tau$ takes into account the variability between the individual RTF relations, enforcing a covariance amongst them. However, note that this does not curtail the use of individual RTF relations to asses the Hubble variance, but aids an opportunity to construct a `standardizable'  sample of galaxies that follow a scaling relation, and can be utilized for additional distance ladder analysis, which we intend to perform as an independent full-fledged analysis. As we shall elaborate later in the \Cref{sec:Joint_analysis}, we do not find strong correlations between individual radial bins, and therefore we leave this assessment to a future analysis. 

To perform the fully Bayesian analysis we utilize the publicly available \texttt{emcee}\footnote{\href{http://dfm.io/emcee/current/}{http://dfm.io/emcee/current/}} package \citep{Foreman-Mackey13}, implementing an affine-invariant ensemble sampler. We analyse the generated MCMC samples using \texttt{corner} and/or \texttt{GetDist} \footnote{\href{https://getdist.readthedocs.io/}{https://getdist.readthedocs.io/}} \cite{Lewis:2019xzd} packages.
\begin{figure}
    \centering
    \includegraphics[scale = 0.45]{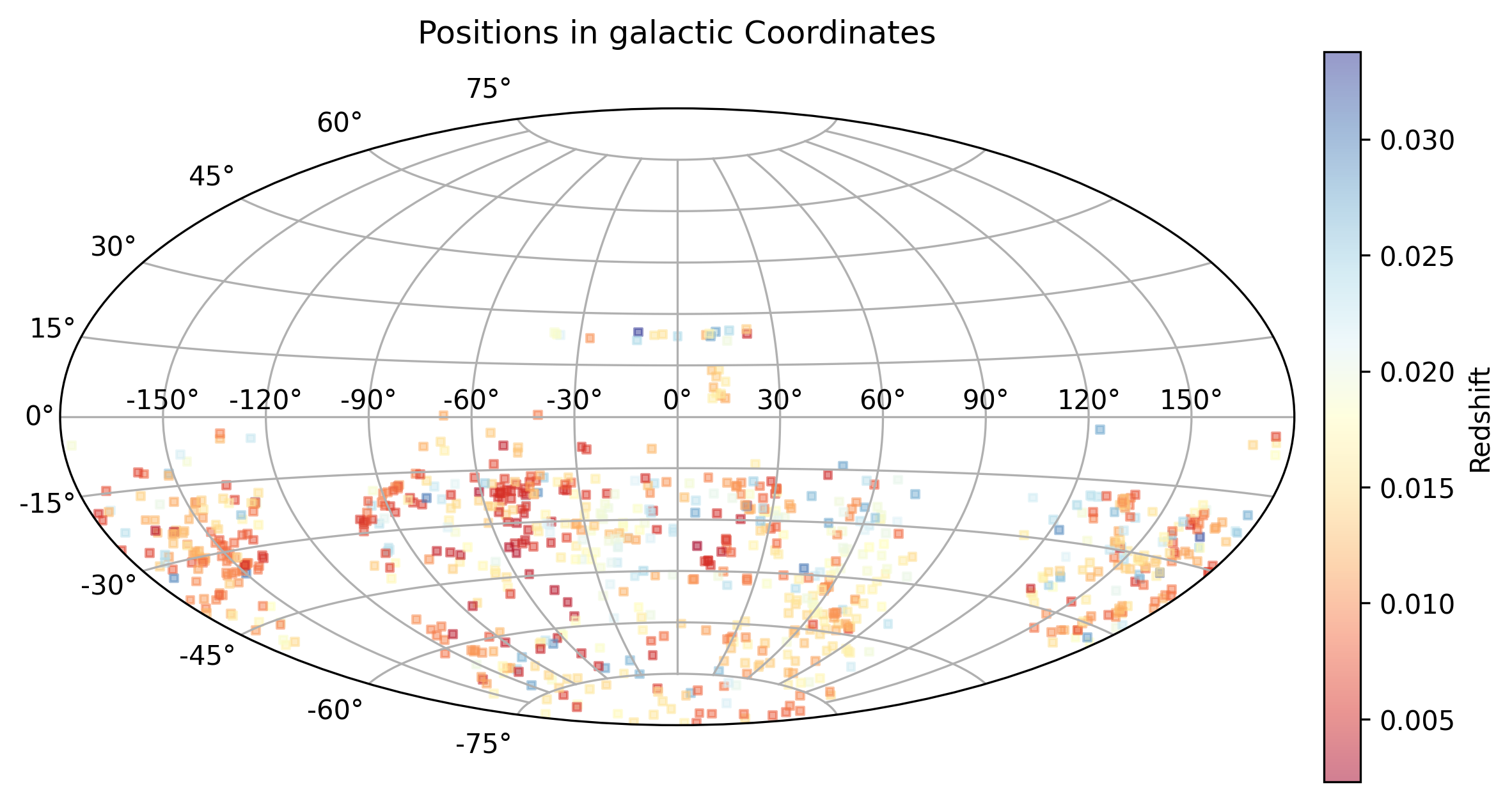}
    \caption{Distribution of the galaxies utilized in the current analysis. The vertical color bar shows the redshift distribution of the galaxies. }
    \label{fig:sky}
\end{figure}

\subsection{Isotropy of the local universe}
Given the availability of the galaxy positions \cite{Mathewson:1992ia}, we also estimate the isotropy and the subsequent variance in the sky of the current sample. Note that the current sample only covers the southern sky and is not an isotropic survey allowing us to assess the overall isotropy being biased on large scales. Therefore, we estimate the overall noise level that would be expected in an isotropic universe, utilizing the bootstrapping methodology presented in \cite{Soltis:2020gpl}. In this method, the positions of the galaxies are permuted amongst themselves essentially rearranging under the assumption that in an isotropic universe, the galaxies could have been observed in any of the given positions within the sky coverage of the survey. In \cite{Soltis:2020gpl}, the residuals of the supernovae magnitudes within the MCMC fitting are utilized as the indicators for the noise levels. Similarly, we utilize

\begin{equation}
    r_{\rm{i}} = \frac{M_{\rm{I, i}}^{\rm{theory}} - M_{\rm I, i}^{\rm data}}{\sigma_{M_{\rm I, i}^{\rm data}}}
\end{equation}
as the indicators for the same, where $i$ iterates over the number of galaxies. The assumption of these residuals as an indicator is valid as we do not intend to estimate the actual isotropy itself but rather the noise level present within the distribution of the galaxies and contrast against the observed positions of the galaxies. Using these residuals in the sky we estimate the clustering of the galaxies by estimating the angular power spectrum $C_{\rm l}$, the range of which depends on the assumption of the NSIDE\footnote{The NSIDE fixes the resolution of the maps. } of the constructed maps. In this analysis, we present our results for NSIDE =16, while we validate that a different assumption does not change our inference substantially. For each of the galaxies, we take the total residual obtained from all the available optical bins and then an average value of these residuals for all the galaxies falling within each pixel of the map. We perform 1000 bootstrap steps for every set of the model parameters within the MCMC chain that are iterated 1000 times, yielding a total of a million steps taking into account uncertainty in the RTF relations. It is important to note that the linear regression of the RTF relations described in the earlier sections is performed assuming no uncertainty ($\sigma_{M_{\rm I, i}^{\rm data}} = 0$)\footnote{The anticipated intrinsic scatter of the RTF relations is much larger than error on the magnitudes of the galaxies and the inclusion of $\sigma_{M_{\rm I, i}^{\rm data}}$ error would fall completely within the prior. For this reason, we remain with fitting eh RTF relations to estimate the maximum scatter.} on the $M_{\rm I, i}^{\rm data}$. However, in estimating the noise level associated with the statistical isotropy, we assume the reasonable conservative value of  $\sigma_{M_{\rm I, i}^{\rm data}} = 0.075$ \citep{Mathewson:1992ia} for the photometric measurements errors.   

\section{Results}
\label{sec:Results}

{\renewcommand{\arraystretch}{1.7}%
    \setlength{\tabcolsep}{5pt}%
    \begin{table*}
        \centering
        \caption{We show the posteriors, including the $68\%$ C.L. limits for the linear regression parameters, performed for each of the radial bins shown in \Cref{fig:PS_95}. The first column is the radial bins and the next three columns present eh results obtained by fitting the RTFs individually. In columns 4 and 5 we show the results obtained in the joint analysis. In the last column, we show the number of data points utilized in the regression in each bin. All constraints reported here are obtained assuming $H_0 = 70 \,\ksM$. }
        \label{tab:LRS}
        
        \begin{tabular}{c|cccccc }
        \hline
        \multirow{2}{*}{$R_n\, [\Ropt]$} & & & &  \multicolumn{2}{c}{Joint} & \multirow{2}{*}{$N_{\rm D}$}\\ \cline{5-6}
        & $b_n$ & $a_n$ & $\sigma_n^{\rm int}$ & $a_n$ & $\sigma_n^{\rm int}$  & \\  
    \hline
    0.2 & $-11.98 \pm 0.11$ & $-4.67 \pm 0.06$ & $0.402 \pm 0.011$ & $ -4.62 \pm 0.04 $ & $ 0.370 \pm 0.012 $ & 749 \\ 
    0.4 & $-8.18 \pm 0.07$ & $-6.15 \pm 0.03$ & $0.191 \pm 0.005$ & $ -6.16 \pm 0.03 $ & $ 0.108 \pm 0.010 $ & 793 \\
    0.6 & $-5.78 \pm 0.06$ & $-7.11 \pm 0.03$ & $0.148 \pm 0.004$ & $ -7.11 \pm 0.02 $ & $ 0.013 \pm 0.009 $ & 799\\
    0.8 & $-4.21 \pm 0.09$ & $-7.72 \pm 0.04$ & $0.173 \pm 0.005$ & $ -7.76 \pm 0.03 $ & $ 0.068 \pm 0.016 $ & 663\\
    1.0 & $-3.02 \pm 0.15$ & $-8.20 \pm 0.07$ & $0.214 \pm 0.007$ & $ -8.18 \pm 0.04 $ & $ 0.145 \pm 0.012 $ & 454\\
    1.2 & $-2.06 \pm 0.26$ & $-8.61 \pm 0.12$ & $0.262 \pm 0.013$ & $ -8.44 \pm 0.06 $ & $ 0.157 \pm 0.004 $ & 231\\
    \hline
    \end{tabular}
    \end{table*}
}

\begin{figure}
    \centering
    \includegraphics[scale = 0.58]{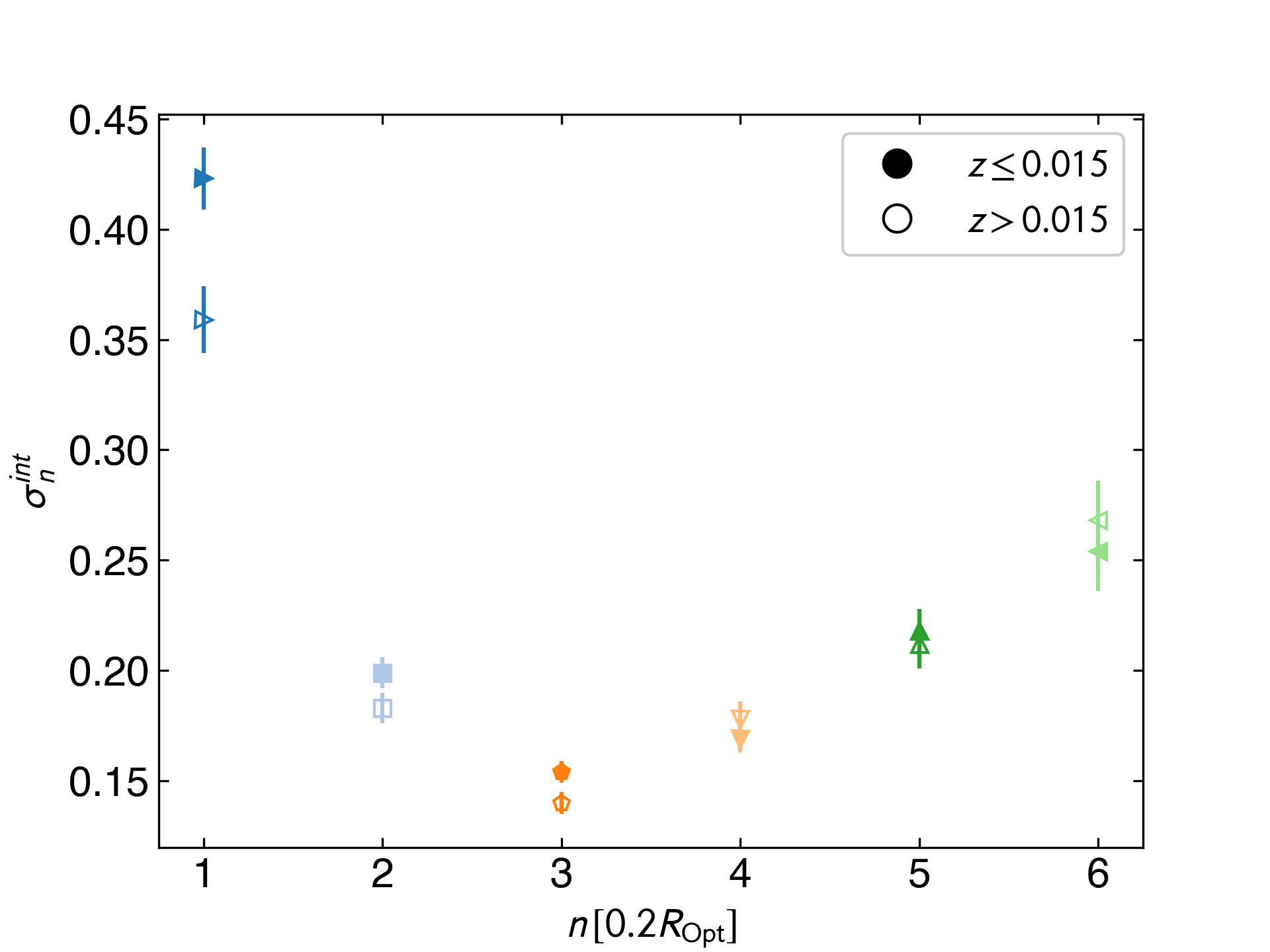}
    \caption{ The intrinsic scatter and the $1\sigma$ error, within each of the radial bins, however, split at redshift $z \sim 0.015$ to assess the variation in the different redshift bins.  }
    \label{fig:Int_scatter_z}
\end{figure}

We begin by re-analyzing the \citetalias{Persic:1995ru} sample containing 6 different radial bins ($\Ropt$). The results obtained for the fitting of the individual RTFs are summarised in \Cref{tab:LRS} for completeness, and are consistent in comparison to those presented in \citetalias{Yegorova:2006wv} (see Table A1. therein). In the current Bayesian formalism the mean values of the scatter are mildly larger along with associated uncertainty. Following this to assess the redshift variation, as a preliminary step, we split the galaxy sample into two redshift bins performing the linear regression in each of the radial optical bins. In \Cref{fig:Int_scatter_z} we show the intrinsic scatter as obtained for each of the radial bins when the regression is performed with only samples $z \leq 0.015$ and $z \geq 0.015$ are considered independently. We find a very good consistency between the intrinsic scatter estimated for the galaxies with the redshift cut which asserts that a redshift-dependent analysis would not be biased due to the variation in the galaxy dataset over redshift. This allows us to estimate the variance of the Hubble parameter as a function of redshift and, consequently luminosity distance. For the innermost optical bin $0.2\, \Ropt$, we find the intrinsic scatter is mildly larger for $z<0.015$, however consistent within $2 \sigma$. Given the large intrinsic scatter in this bin, mostly due to the random presence of a not-negligible compact bulge component in our galaxies, it is difficult to estimate the rotational velocity very accurately along the lines of the bulge-free \Cref{eqn:Vmodel}. Therefore, we exclude this bin in our baseline analysis when estimating the overall variance \footnote{ The effect of the bulge in RTF is discussed in \citetalias{Yegorova:2006wv}.}. We also exclude the outermost bin, owing to the low sample density having only $\sim 230$ data points, and the fact that the accuracy in measuring the rotational velocity from the $H_{\alpha}$ line is low at the outskirts of the spiral galaxies. Nevertheless, we retain the advantage of utilizing 4 independent radial bins to estimate the overall scatter. From here onwards we remain with 4 optical bins as our `baseline' dataset to present our main results to evaluate the cosmological variance of the Hubble parameter. However, we do comment on the implications of utilizing these two bins as they present interesting scenarios in assessing the Hubble variance. The $0.6\, \Ropt$ bin having the least scatter\footnote{Also for the RTFs emerging in the \cite{Vogt2004, Courteau1997} samples.} and being best constrained RTF, is a conservative estimate to which we compare the joint constraint in our analysis.

\begin{figure}
    \centering
    \includegraphics[scale = 0.58]{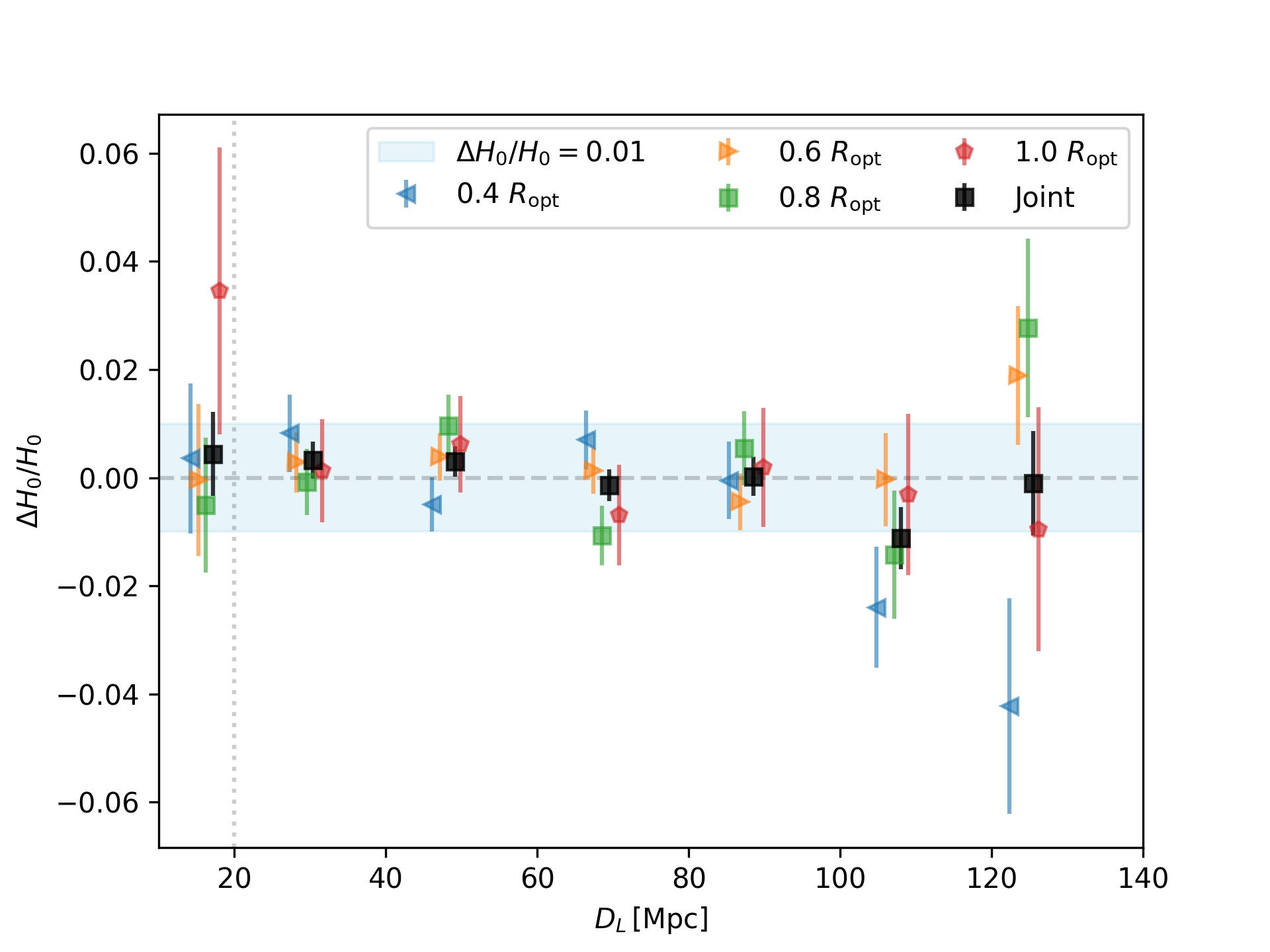}
    \caption{The variation in the Hubble parameter utilizing all 5 radial optical bins simultaneously. The data points have been slightly shifted in the x-axis for better visualization.}
    \label{fig:Hubble}
\end{figure}

Having the Tully-Fisher relation fitted in each of the bins, and establishing that the dataset is suitable for the assessment of redshift variation, we now evaluate the residual of the absolute magnitude, w.r.t the fitted linear regression. We conservatively anticipate the intrinsic scatter and hence the absolute residual, $\Delta \Mi$ to account for cosmological variation which can be converted to the variance in the expansion rate as given by \Cref{eqn:cosmo_variation}, also considering the uncertainty of the RTF relations\footnote{We validate that including the uncertainty of the RTF relation itself makes very little difference to our final estimates, since the intrinsic scatter much larger in comparison.}. Note that we also exclude the galaxies that are more that more than $3\sigma$ away from the residuals deeming them to be outliers\footnote{We find only about 5-10 galaxies per radial bin to satisfy the outlier condition, which however could be an important assessment as the lower $\DL< 20\, [\Mpc]$ and $\DL> 120 \, [\Mpc]$ distance bins with a low number of galaxies can get mildly affected. We also validate that this step does not make a significant difference to our final inference. }. Following \Cref{eqn:Cosmography_final}, we now translate the average dispersion in the residual of RTF relation as the maximum possible variance in the value of $\Hzero$ as, 
\begin{eqnarray}
\label{eqn:cosmo_variation}
    \frac{\Delta \Hzero}{\Hzero}(z) &=& \frac{\ln{10}}{5}\overline{\Delta M}(z).
\end{eqnarray}

Which is now recast as the redshift evolution of the allowed fractional variation in the Hubble parameter as shown in \Cref{fig:Hubble}, by binning the residuals appropriately. In this figure, we show the mean of the residuals for the best-fit RTF relations and the uncertainity on the same, estimated as the standard error. We present the residuals in all the radial bins while assessing the corresponding cosmological distance ($\DL$) using \Cref{eqn:Cosmography_final}, assuming that the redshift is given by the recession velocity ($\Vrec$) as $z = \Vrec/c$. We then compute the average variation of the $\Delta \Mi$ in each of the redshift/distance bins split accordingly. Needless to say, the difference in the number of data points (galaxies) in each bin is reflected in the uncertainity estimated as standard error $\sigma_{\rm \overline{\Delta \Hzero/\Hzero}} = \sigma_{\rm{\Delta \Hzero/\Hzero}}/\sqrt{n_{\rm{i}}}$, where $n_{\rm{i}}$ is the number of data points (galaxies) in each redshift bin. As we have already mentioned, each of the RTF relations fitted in the 6 optical bins can be considered independent relations with no correlations whatsoever. Therefore, in the joint analysis of all the radial bins $R_n [\Ropt]$ considered, we follow the same procedure considering each velocity measurement in each radial bin as an independent data point. In \Cref{fig:Hubble} we show the variation in the residuals of absolute magnitudes of the galaxies, as a function of the luminosity distance, wherein we have binned the luminosity distance into linear bins of size $\Delta D_{\rm L} = 20 \, [\Mpc]$. 

As can be seen in \Cref{fig:Hubble}, we find that there is no discernible redshift evolution of the Hubble variance, always consistent within $\sim 5 \%$, suggesting no variation in the current redshift range. We find that the joint constraint to sub percent precision always being consistent with $\DHzero = 0 $ within the $\sigma_{\rm \overline{\Delta \Hzero/\Hzero}}$. For the $0.4\, \Ropt$ we find the maximum variation to be $\sim 10\%$ comparing the data point at $\sim 120 \, \Mpc$ with the innermost data point at $\sim 17\, \Mpc$. Similarly, the $1.0\, \Ropt$ also shows a mild increase in the variance in the innermost distance bin. However, this trend is not immediately corroborated in the $0.6\, \Ropt$ and $0.8\, \Ropt$ radial bins, which are the better-constrained RTF relations showing no statistically significant trend. This is also reflected in the joint constraint shown as black data points in \Cref{fig:Hubble}. Moreover, there seems to be a mild decrease in the values of $\Delta \Hzero/ \Hzero$, especially around the $D_{\rm L} = 100\, [\Mpc]$ range, however, completely consistent within the $\sim 2\sigma$ C.L. for $\DHzero = 0$. 

We contrast our results for the uncertainty in the evolution of the Hubble parameter against the constraints obtained from the Supernovae datasets in the local universe. Although the current galaxy dataset utilized in this work only extends up to $D_{\rm L} \sim 140 [\Mpc]$, we confirm that within this range such an under-density is not suggested by utilizing the RTF relation. This is in agreement with the earlier analysis in \cite{Kenworthy_2019, Lukovic19, Camarena19} using SNe datasets. An under density of size $\sim 300 \, [\Mpc]$ \cite{Hoscheit17} and a similar less significant local hole of $\sim 150\, [\Mpc]$ \cite{Shanks19, Whitbourn14} were proposed as possibilities to alleviate the Hubble tension by modifying the current expansion rate measured by the local supernovae. To comment on which one would need to extend the current sample to a higher redshift range \cite{Stone2022} (left for a future investigation). Within the redshift range of the current dataset, we find that local 
$ z< 0.01$ sharp phantom transitions as proposed in \cite{Alestas:2020zol} are less likely. 
Similarly, sharp transitions in the gravitational constant ($G_{\rm N}$) local/ultra-local universe \cite{Marra:2021fvf, Alestas:2021luu, Benisty:2023laj, Camarena:2023rsd, Perivolaropoulos:2023iqj} and local modifications \cite{Ruchika:2023ugh} have also been suggested as a possibility for the discrepancy in the locally measured $\Hzero$ and the value inferred from the CMB assuming model, which our results so far do not immediately suggest. In other words, we notice that in the current linear binning of distance, the innermost constraint obtained at $\DL\sim 17 \, [\Mpc]$ is completely consistent with no evolution of the $\DHzero$. This constraint is also consistent with the second bin centered at $\DL \sim 30\, [\Mpc]$. As can be seen in \Cref{fig:Hubble}, our conservative inference using only the radial bin $0.6 \, \Ropt$ is completely consistent with the joint constraint of the 4 considered so far. However, please refer also to the \Cref{sec:Im_bin,sec:logbin} for mild yet alternate perspectives mostly arising due to different binning schemes and the inclusion of inner and outermost radial bins. 

\subsection{Inclusion of the innermost bin and alternate binning}
\label{sec:Im_bin}

\begin{figure}
    \centering
    \includegraphics[scale=0.57]{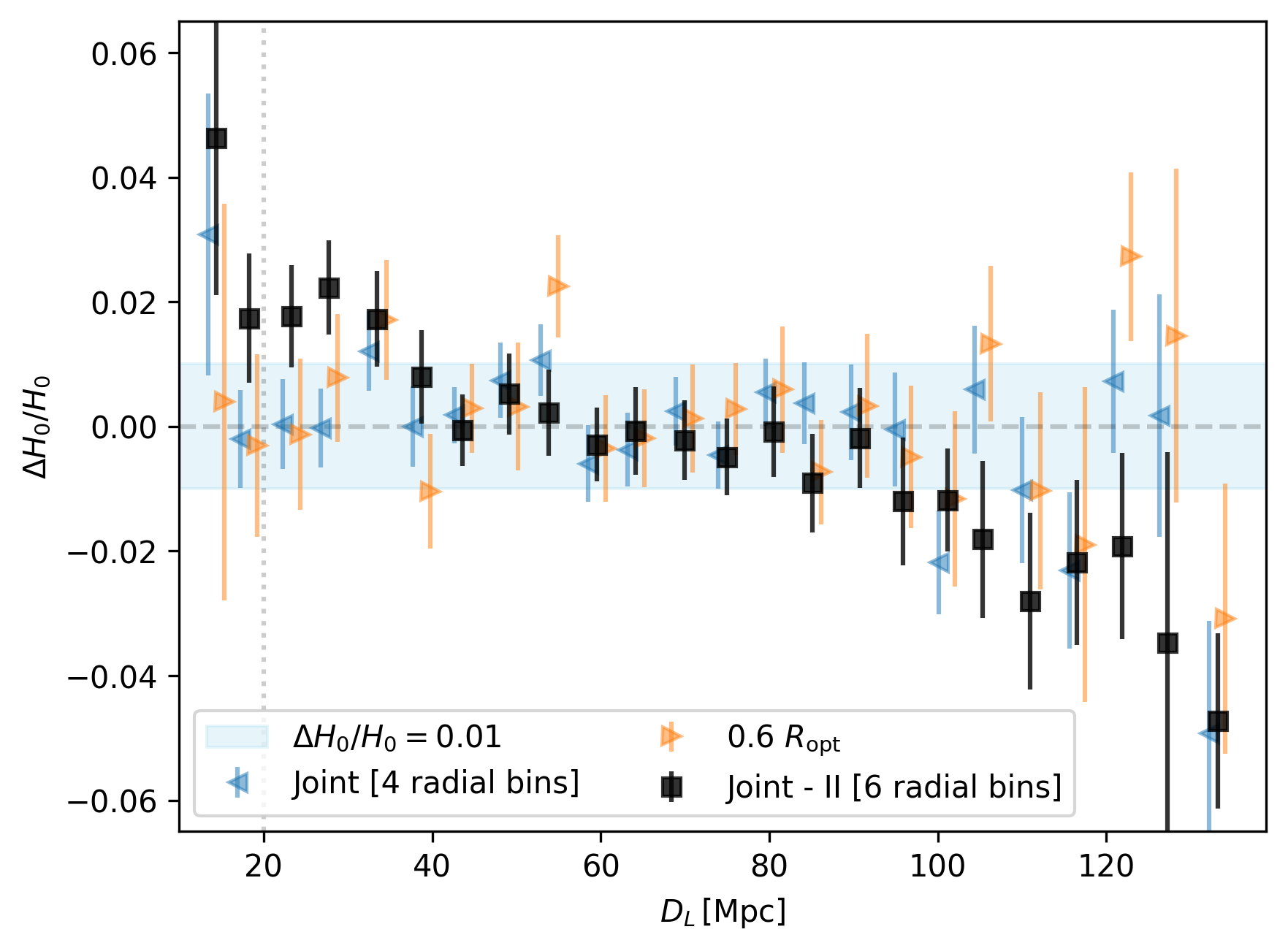}
    \caption{Same as \Cref{fig:Hubble} with different binning schemes and inclusion of the innermost radial $0.2\, \Ropt$ bin. Black data points show the results using the same binning as in \Cref{fig:Hubble}.}
    \label{fig:Hubble_6}
\end{figure}

In this section, we present the results as obtained when the innermost $0.2\, \Ropt$ and outer bins $1.2\, \Ropt$ are also included in the analysis. We also change the binning scheme of the distances to test for the validity of including the innermost bin while accommodating variations also on the closet and the farther-distance bins. In \Cref{fig:Hubble_6}, the black data points show the joint constraint taking into account all the $\Ropt$ bins and increasing the number of distance bins in contrast to the earlier results in \Cref{fig:Hubble}. As one can immediately notice the the joint constraints show a trend of increasing $\DHzero$ for lower distances. This change is completely driven by the $0.2 \, \Ropt$ bin alone, while the outermost bin, having only $231$ data points,  makes minimal difference to joint constraints. 

We show results for finer binning in distances with $\Delta \DL \sim 5 \, [\Mpc]$, at the expense of precision in each bin to evaluate the overall shape of $\DHzero(z)$. We find that the $0.2 \, \Ropt$ strongly influences the joint constraint, showing significant evidence for a possible cosmological signature for an increasing value of the Hubble parameter in the `ultra-local' ($\DL \lesssim 30 \, [\Mpc]$) regime. Needless to say, to make this inference is heavily reliant on the binning scheme, as the closer distance bins tend to have very few galaxies providing the jump we notice therein. For instance, within the joint analysis of all the radial bins (black data points in \Cref{fig:Hubble_6}), the first bin centered at $\DL \sim 14.4 \, [\Mpc]$  and the last bin at $\DL \sim 133.2 \, [\Mpc]$ contain merely 5 and 3,  galaxies, respectively. In this case the difference in the bins is $\DHzero \sim 0.1$, which is about $\sim 10\%$ variation in the value of absolute $H_0$. Note that this level of variation difference could play a significant role in explaining the $H_0$-tension which is $\sim 8 \%$ difference between the local and the CMB estimates. We also show and validate that the $0.6\,\Ropt$ radial bin shows no such behavior using the finer bins, with mildly increasing scatter of the mean at larger distances. 

The initial consideration to leave this bin out of the joint analysis is because the central bulge in the spiral galaxies does not allow one to measure the rotational velocity accurately. Therefore, the validity of the redshift-dependent behavior of $\DHzero$ we find in \Cref{fig:Hubble_6}, is subject to the confidence in the rotational velocities measured in the innermost regions of the spiral galaxies. In \citetalias{Yegorova:2006wv}, the authors provide a relationship (see Eq.14 therein) between the slopes of the individual RTFs and the optical radii, indicating the validity of the RTFs also in the bulge of the galaxies. This is once again an empirical relation that is observed and fitted in the current galaxy sample and it is also shown that this expectation is consistent when considering the two other galaxy samples they have tested. However, in a conservative approach, we tend to remain with our 4-bin analysis or equivalently 5 bins including the outermost radial bin, not claiming significant evidence for a redshift evolution of $\DHzero$. This variation has to be validated in larger galaxy samples with better estimations of the central rotational velocities and would be untimely to claim a possible detection of local variation in $H_0$ using the current dataset alone. We leave this for a future investigation using the PROBES \citep{Arora2021} dataset. Howbeit, it is interesting to note that the innermost bin carries significant statistics being able to sway the joint analysis using all 6 bins. 

\begin{figure}
    \centering
    \includegraphics[scale = 0.58]{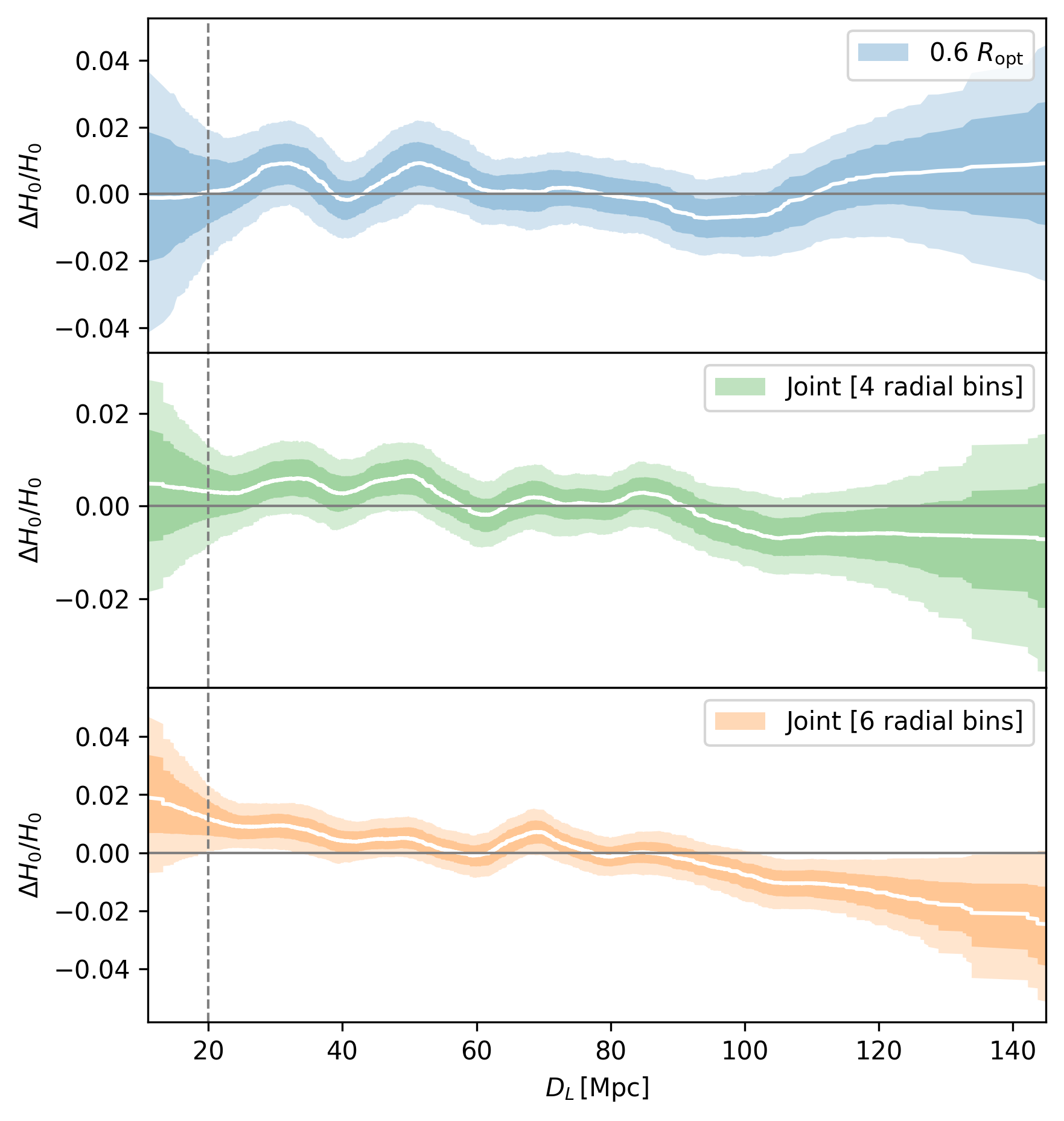}
    \caption{\textit{Top}: We show the $1\sigma $ and $2\sigma$ C.L. of the  fit to the $0.6 \Ropt$ bin. \textit{Center}: Similar to the top panel, for the baseline Joint analysis with the 4 radial bins and for comparison, we also show the results from all the 6 radial bins in the \textit{Bottom} panel. }
    \label{fig:LOESS}
\end{figure}

\subsection{Smoothing the binned variance}
\label{sec:LOESS}

While we have so far presented our results by binning the galaxies in distance, we now implement a simple smoothing of the scatter using the locally weighted scatter regression of the scatter (LOWESS \footnote{Locally weighted scatter plot smoothing (LOWESS) \citep{Cleveland1988Sep, Cleveland1979Dec}. We utilize the publicly available \texttt{statsmodels} package \cite{seabold2010statsmodels} for this purpose. A simple example of how to perform the same can be found  \href{https://www.statsmodels.org/stable/examples/notebooks/generated/lowess.html}{here}. Please see also \cite{Cook1999Jul, Fox2008, Fox2015, Andersen2009Jul} for more details.}) obtained in the $\DHzero \, \rm{vs.}$ $ \DL$ plane. This technique has been utilized in various contexts in \cite{Montiel:2014fpa, Escamilla-Rivera:2021rbe, Bernardo:2021mfs, Fernandez-Hernandez2019Oct}. We implement this essentially to represent the variation in $\DHzero$ as a smooth function of $\DL$ and simultaneously obtain the uncertainty on the estimated variance. Similar to the binning schemes where we have to assume the size of the $\Delta \DL$, here we need to assume a fraction ($f\in \{0,1\}$) of the data that will be considered to obtain the locally weighted least squares fit. Higher values of this fraction will consider a larger number of data points eventually providing a smoother reconstruction of the scatter plot. This technique while being 'non-parametric' also allows one to obtain the uncertainty region by simply bootstrapping on the scatter points. We utilize this method to present our final constraints on the $\DHzero$ as our main result in terms of uncertainty and the upper limits.

We show the results of the LOWESS smoothing in \Cref{fig:LOESS} for three different scenarios: using only the central $0.6 \, \Ropt$ bin alone, our baseline analysis with 4 radial bins and all 6 radial bins included. Here we have utilized $f = 0.2$, while we have tested the results also with the larger fractions. We show a comparison of the constraints and $95\%$ upper limits on $\DHzero$ in \Cref{tab:loess}, for $f = 0.2$ and $f= 0.5$ fractions. Using lower fractions ($f< 0.1$) tends to provide very few galaxies for each locally weighted regression, essentially mimicking the scatter with very large mean squared errors. Note that traditionally the fraction of data to be utilized is optimized for the least possible mean square error through cross-validation techniques \cite[see references therin]{Montiel:2014fpa}. In our case, this typically corresponds to the larger values of $f$, as can be seen in the last column of \Cref{tab:loess}. Therefore we remain with the conservative $f =0.2$ fraction of data to present the limits expected on $\DHzero$. 

As shown in \Cref{tab:loess}, using only the $0.6\, \Ropt$ bin we find $\DHzero\, [\%] = -0.11^{+2.06}_{-1.96}$ with a $95\%$ C.L. upper limit of $\DHzero < 3.9 \%$. We interpret this limit as a conservative upper limit on the allowed local variation on $\Hzero$. Note that this limit is estimated at a distance $\DL \sim 10 \, [\Mpc]$. Similarly, we find the upper limit at $95 \%$ C.L. limits is $\DHzero \lesssim 3 \%$ which we infer as our best estimate. Needless to say, these limits are tighter when we utilize $f =0.5$, as can be seen in the last two columns of \Cref{tab:loess}. We also find that similar features are reconstructed in both these cases, validating the use of 4 bins as our baseline dataset. We notice a mild dip at $\DL \sim 100 \, [\Mpc]$ and a rise at $\DL \sim 30 \, [\Mpc]$, albeit with a low significance of $\sim 1\sigma$. Finally, with the inclusion of the innermost radial bin, the redshift dependence of $\DHzero$ is evident as depicted in the bottom panel of \Cref{fig:LOESS}. This also translates to a detection of $\DHzero > 0$ at C.L. of $ \sim 1 \sigma (f = 0.2)$ and $\sim 2\sigma (f = 0.5)$.

{\renewcommand{\arraystretch}{1.8}
\setlength{\tabcolsep}{3pt}

\begin{table}
\centering
\begin{tabular}{c|c c c c}
\hline
\multirow{3}{*}{Bins} & \multicolumn{4}{c}{$\DHzero [\%]$} \\
 & \multicolumn{2}{c}{$f = 0.2$} & \multicolumn{2}{c}{$f = 0.5$} \\
 & $68\% $ C.L. & $< 95\% $ C.L. & $68\% $ C.L. & $< 95\% $ C.L. \\ 
\hline

$0.6 \Ropt$ & $-0.11^{+2.06}_{-1.96}$& $3.90$ & $0.35^{+1.04}_{-1.18}$ & $2.64$ \\

4 bins & $0.54^{+1.32}_{-1.37}$ & $2.98$ & $0.34^{+0.63}_{-0.74}$& $1.74$ \\

6 bins  & $2.08^{+1.65}_{-1.39}$ & $5.17$ & $1.45^{+0.72}_{-0.69}$ & $2.79$ \\
\hline
\end{tabular}
\caption{We show the $95\%$ C.L. upper limits on the $\DHzero [\%]$, for the two cases of the fraction of data utilized in the analysis. The $f=0.2$ column corresponds to the uncertainty regions presented in \Cref{fig:LOESS}.} 
\label{tab:loess}
\end{table}
}


\subsection{Joint analysis}
\label{sec:Joint_analysis}

\begin{figure}
    \centering
    \includegraphics[scale=0.58]{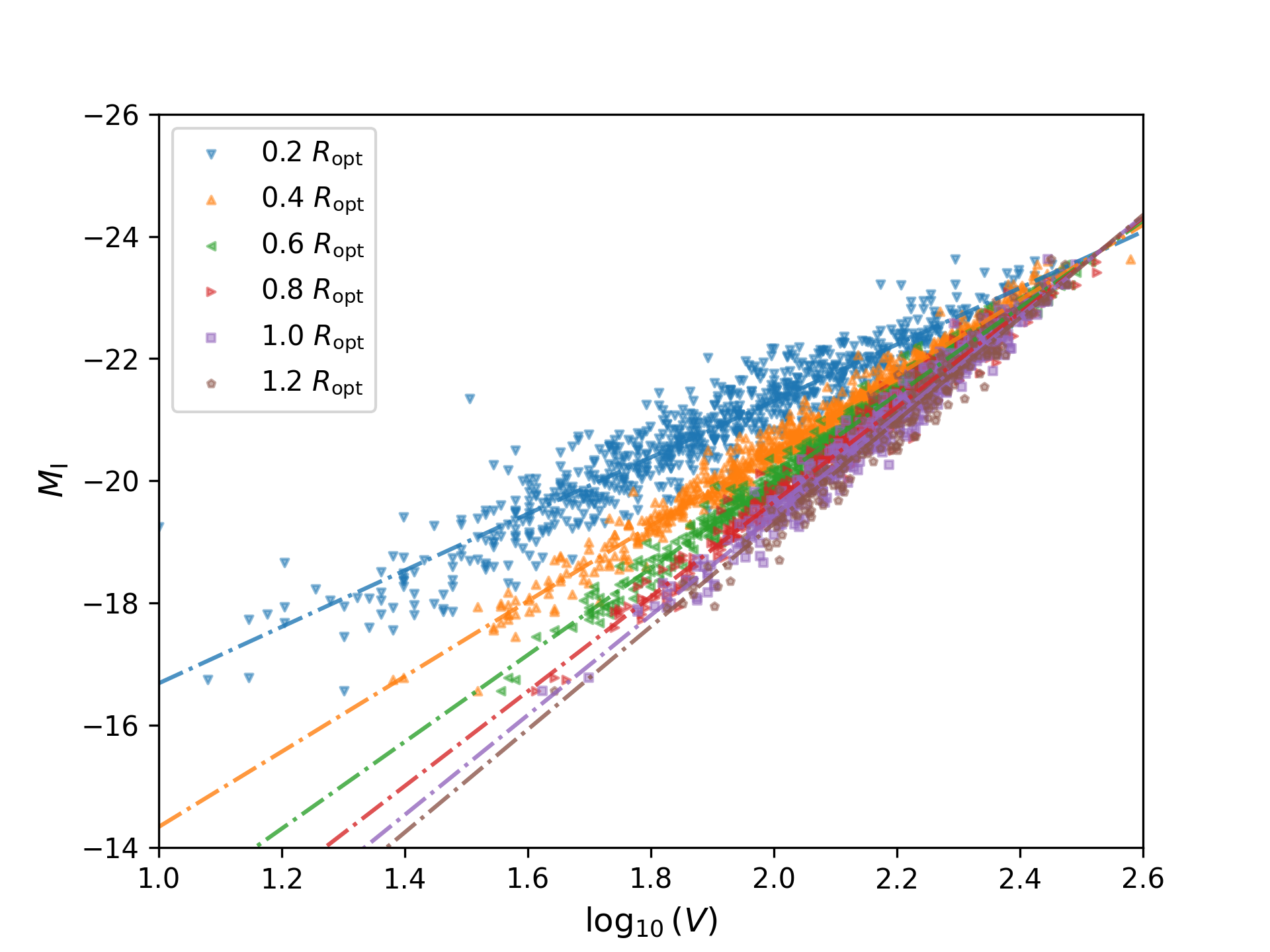}
    \caption{Scatter corresponding to the RTF relations in the 6 bins of $\Ropt$ taken from \citetalias{Persic:1995ru} and the linear best-fits obtained from our joint analysis assuming $\Hzero = 70 \, \ksM$.}
    \label{fig:PS_95}
\end{figure} 

As anticipated in \Cref{sec:method}, we perform the joint analysis of the RTFs considering the pivot point as parameters of the fit and individual slopes and intrinsic scatters for each of the RTFs as free parameters. To validate this assumption, we first perform a rolling regression\footnote{We use \texttt{PyMC} \cite{pymc,pymc3} for this purpose.} analysis where no binning is considered but all the data points from all the bins are simultaneously fitted assuming a gradually changing slope. Wherein we recover that all the regression lines pass through the pivot point, without an explicit assumption of the same. In \Cref{fig:PS_95}, we show the RTF relations fitted against data in this approach. As can be seen in \Cref{fig:PS_95} and also in \Cref{tab:LRS} we find a very good agreement with the scenario when performing individual fits, especially contrasting the slope of the RTF relations. We obtain the following constraints for the pivot parameters,

\begin{eqnarray*}
    \MI &=& -23.830 \pm 0.066,\\
    \logV &=& 2.526 \pm 0.009. \\
\end{eqnarray*}

The estimates of the intrinsic scatter and their uncertainty, are at times lower than those obtained in the individual fits. This is clearly because a fraction of the uncertainty is now attributed to the estimation of the pivot, and reduction in the overall degrees of freedom, in the joint analysis. In \Cref{fig:Corr}, we show the correlations between the posteriors of the slopes and the pivot point. As expected there exists a strong anti-correlation between the parameters $\logV$ and $\MI$. The slopes of the individual RTF relations are correlated with the coordinates of the pivot point, ranging from anti-correlations for the inner-most $0.2\Ropt$ RTF to positive correlation for the outer-most $1.2\Ropt$ RTF. In this context, it is interesting to note that the slope of $0.6\Ropt$ RTF is almost completely uncorrelated to the pivot point and shows only mild correlations with the slopes of the other RTFs. This also validates the existence of a very well-constrained individual RTF relation in this optical bin.

We show the complete contour plot of all the parameters of the MCMC analysis in \Cref{fig:corner_full}, for brevity in the main text. The analysis so far has been performed assuming $H_0 = 70\,\ksM$, which is necessary to estimate the distances to galaxies. In \Cref{fig:H0-comp} we show the comparison the posteriors when the $H_0$ is assumed differently, being either $67 \, \ksM$ \cite{Planck18_parameters} or $73 \, \ksM$ \cite{Riess:2020fzl}. As anticipated the magnitude of the pivot point is strongly affected by the assumption, while the velocity remains completely unchanged. Also indicating that not assuming a particular value of $H_0$ mainly affects the overall scale of the RTF relations and not necessarily the shape of individual relations. However, the redshift dependence of the data can be affected by the assumption of underlying cosmology and tentatively this in turn can allow one to use the current dataset to constrain the value of $H_0$. We intend to present this in the second installment of this pilot analysis, utilizing also the necessary local distance calibrators.  

\subsection{Variance on the sky}
\label{sec:iso}
As yet we have only estimated the variation of the Hubble parameter as a function of redshift, however now we turn to the anisotropy in the estimate of the $H_0$ on the sky. For this purpose, we utilize the method presented in \cite{Soltis:2020gpl}, wherein the angular clustering of the Supernovae sample \cite{Riess17_pantheon, Scolnic:2017caz} was estimated and a percent level spatial variation in the Hubble constant was reported. Following this methodology, we compute the spatial clustering of the galaxies in the current \citetalias{Persic:1995ru} sample, which is a collection of galaxies in the southern hemisphere as shown in \Cref{fig:sky}. Note however, that in contrast to the SNe datasets that extend all the to $z\lesssim 2.3$, our dataset of local galaxies is restricted to $z\leq 0.03$, allowing us to test explicitly for the local anisotropy. 

\begin{figure}
    \centering
    \includegraphics[scale = 0.6]{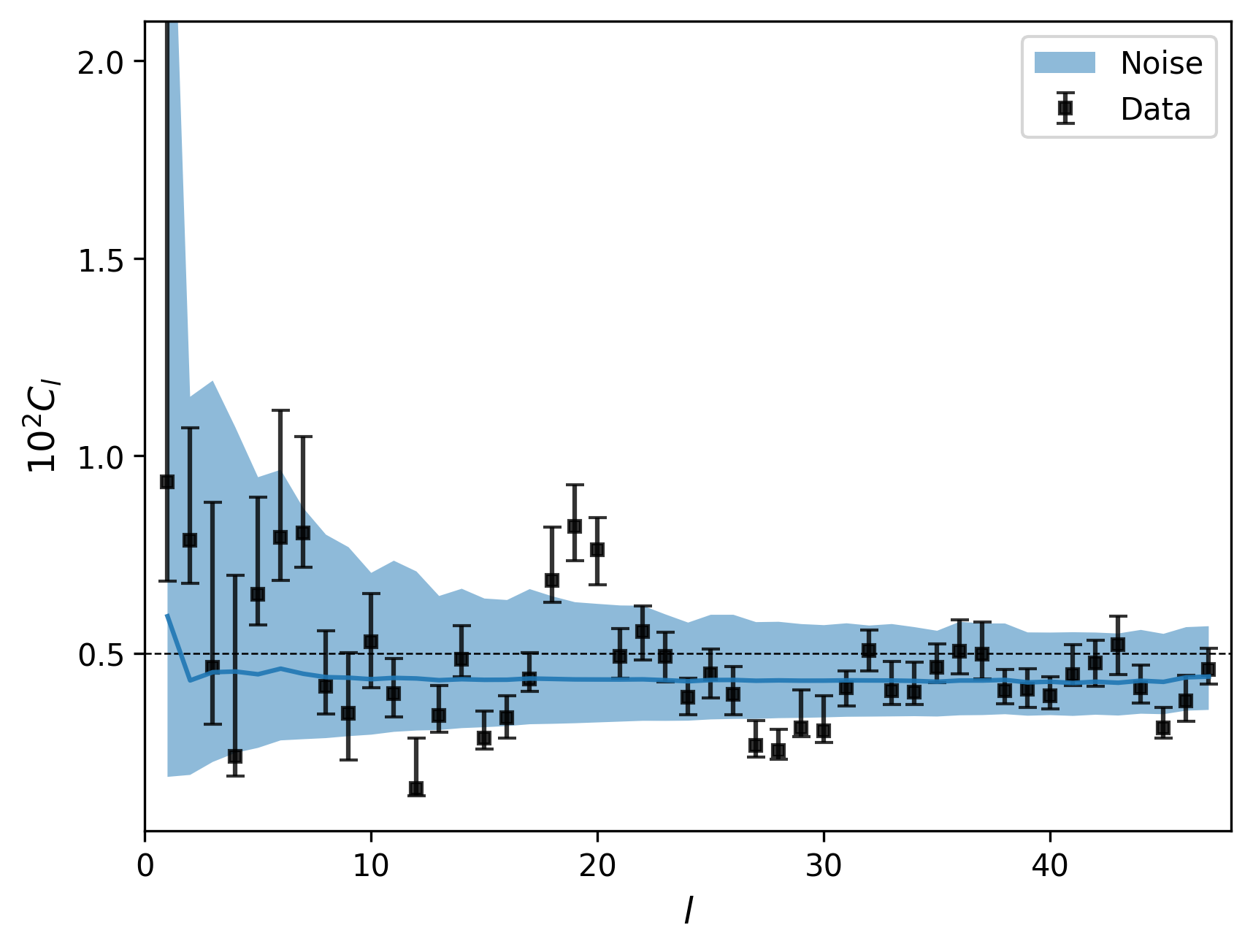}
    \caption{Power spectrum of the spatial clustering of the galaxies. The blue line shows the mean of the noise level, while the shaded region shows the $68\%$ C.L. uncertainty on the noise level. }
    \label{fig:cls}
\end{figure}

In \Cref{fig:cls}, we show the power spectrum of the spatial clustering of the galaxies. The black data points and the associated uncertainties represent the clustering of the galaxies varying on the parameters of the RTF relations as fitted in the MCMC analysis \Cref{fig:corner_full}. The blue-shaded region shows the $68\%$ C.L. uncertainty on the noise level associated with the statistical variation expected in an isotropic universe, given the observed positions of the current galaxies. As can be seen the angular power spectrum of distribution of the galaxies in the current catalog is perfectly consistent within the $\lesssim 1\sigma$ level, with the isotropic expectation, indicating no signal for anisotropy. The current sample only occupies the southern hemisphere, which is reflected in the data and the noise level as large values of the angular power spectrum for $l =1, 2$ and is completely consistent with what is expected. Also, note that the uncertainty in the data shown as the error bars in \Cref{fig:cls}, only takes into account the variation of parameters of the RTFs and a fixed $H_0 = 70 \, \ksM$. We do not anticipate that the values of $H_0$ (also $q_0$) or so any variation in cosmology would affect distances to all the galaxies almost equivalently unless neglecting the redshift dependence in the current redshift range. Therefore, a full assessment can be warranted when $H_0$ is utilized as a free parameter alongside the local distance calibrators, to obtain more quantitative limits on the level of isotropy. This however is not expected to change the inference here that the current galaxy sample is completely consistent with the null hypothesis of isotropy. 

\section{Conclusions and Outlook}
\label{sec:Conclusions}

We have presented the usefulness of the astrophysical scaling relation, in particular, the Radial Tully-Fisher \cite{Yegorova:2006wv}, to assess the variation in the Hubble parameter in the local universe. While similar studies have been performed using the TF relation earlier, here we use RTF for the very first time. This empirical relation exploits the information of the full RC rather than one single reference value as done using the TF relation. From a physical point of view, in contrast to the TF relation, the RTF takes into account the presence of DM in galaxies and the variation in them of the stellar mass-to-light ratios. The primary results of our analysis are summarised as follows: 
\begin{itemize}
    \item We constrain the maximum possible variance in the Hubble parameter as a function of redshift in the range of $0.005< z <0.035$ to be $\DHzero \lesssim 3\%$ at $95\%$ C.L., showing no significant redshift dependence. 
    \item Conservatively, using only the $0.6 \Ropt$ radial bin we find the consistent with our baseline analysis using 4 radial bins with $\DHzero \lesssim 4\%$ at $95\%$, estimated at $\DL \sim 10\, [\Mpc]$.
    \item These constraints allow us to conservatively conclude that any local solutions to alleviate the $H_0$-tension are not supported within the redshift range of the current galaxy sample. 
    \item We introduce a joint analysis of the `independent' RTF relations, reducing the number of parameters while assessing the possible correlation between them. This also provides us with a possible pseudo-standardization of the RTFs.
    \item While occupying only the southern hemisphere, spatial clustering of the current galaxies shows no deviation from the null hypothesis of isotropy.  
\end{itemize}

Constraining the value of the present expansion rate ($H_0$) is one of the most crucial aspects of the current cosmological crisis. Exploring different independent methods and possible synergies in these datasets will yield a better understanding of changing the current cosmological paradigm in a more consistent direction. Needless to say, it is important to validate the analysis with newer datasets. In this context, we intend to extend the current analysis to much larger samples of galaxies, like PROBES \citep{Stone2022} and MANGA \citep{Arora2021}, also extending to farther redshifts \citep{2021MNRAS.503.1753S, 2022A&A...659A..40S} to provide a comprehensive understanding of the evolution of the local and late universe. This is necessary also to validate the possible alternative inference when the innermost radial bin $0.2\,\Ropt$ is included in the joint analysis, which indicates a variation of $\DHzero \sim 10\%$, which is sufficiently apt to explain the $H_0$-tension. 

On the other hand, as a second installment to this pilot study, we intend to perform a joint analysis utilizing the local distance estimators \citep{Riess:2020fzl} to calibrate the RTFs and constrain the value $H_0$. To this end, we have introduced here the joint analysis of the RTFs that will be a preliminary step in this direction, allowing a pseudo-standardization of the empirical Radial Tully-Fisher relations. 

\section*{Acknowledgements}
The authors are grateful to Leandros Perivolaropoulos, Stephane Courteau, and Anto I. Lonappan for insightful discussions. BSH is supported by the INFN INDARK grant.

\section*{Data availability}
The data products utilized in the manuscript are all publicly available and appropriately referenced. And code utilized to perform the analysis here can be made available on reasonable request soon after acceptance of the paper. 


\bibliographystyle{mnras}
\bibliography{bibliografia} 
\newpage


\appendix

\section{Alternate binning of the distance}
\label{sec:logbin}
\begin{figure}
    \centering
    \includegraphics[scale = 0.57]{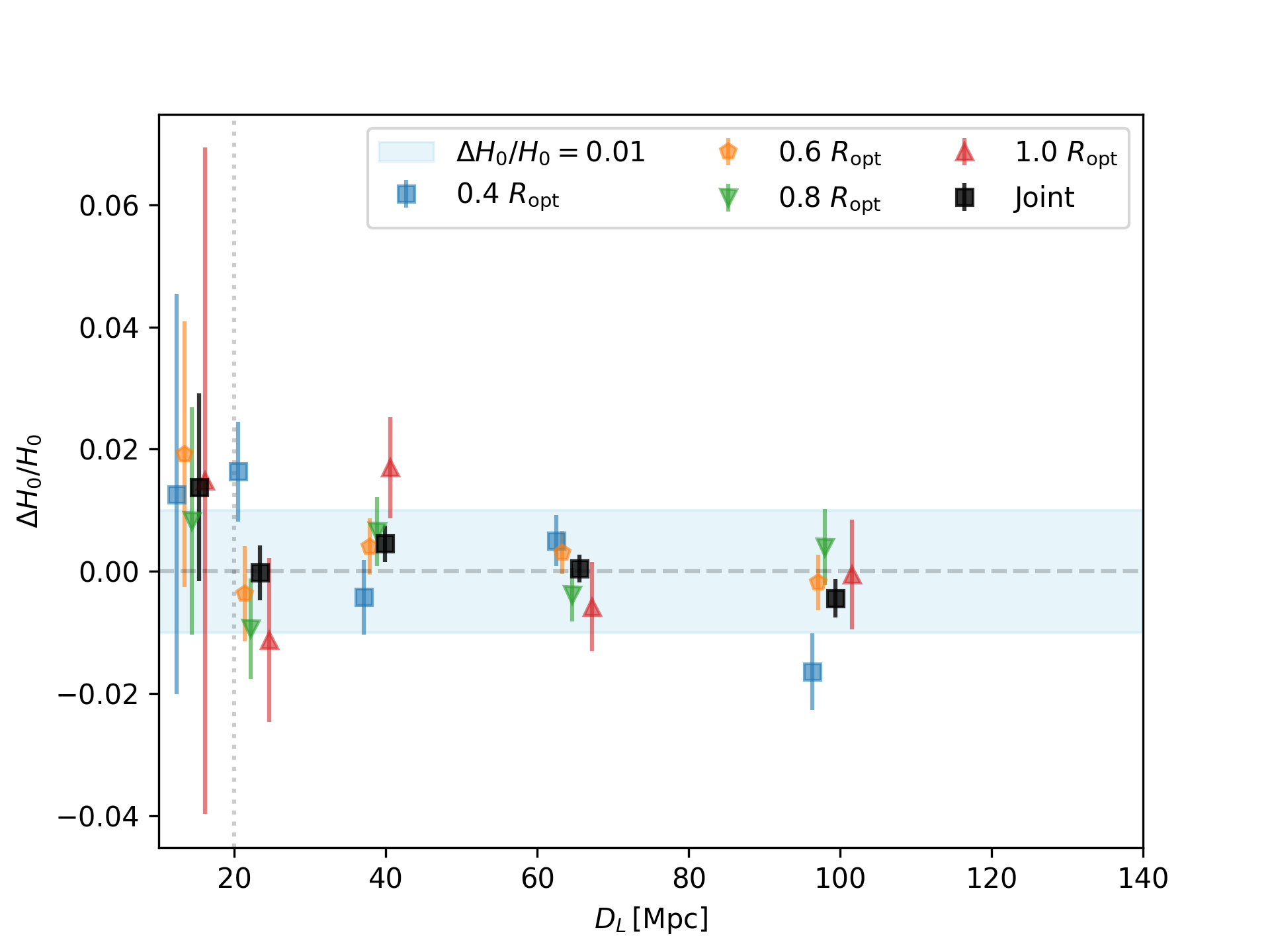}
    \caption{Same as \Cref{fig:Hubble} for the logarithmic binning of distance as described in \Cref{sec:logbin}. }
    \label{fig:Hubble_log}
\end{figure}

In the main text, we have presented our primary results using a bin size of $\Delta D_{\rm L} = 20 \, [\Mpc]$, using linear binning of the magnitude residuals. In this section, we briefly report the mild differences we notice when different binning schemes are utilized. 
In \Cref{fig:Hubble_log} we show the $\DHzero$ using a logarithmic binning scheme. We notice that the joint constraint shows a mild jump in the mean value of $\DHzero$ towards lower redshifts in the closest distance bin, although with no strong significance. And while being completely consistent with $\DHzero = 0$ for distances larger than $20\, [\Mpc]$. Note that this is particularly interesting in the context of some of the ultra late-time resolutions to $H_0$-tension, as we mentioned in the main text. The closet distance bins centered at $\sim 15\,[\Mpc]$ and $\sim 23 \, [\Mpc]$ contain about 10, 100 galaxies in each, respectively. The most local distance bin, with very few galaxies, seems to consistently suggest a mild jump in $H_0$, even for our conservative $0.6 \Ropt$ bin. This as we have already mentioned in the main text is more evident when the innermost $0.2\, \Ropt$ bin is included in the joint analysis. 

\section{Contour plots of the Joint analysis}
\label{sec:con_JA}
For the sake of brevity in the main text, we show the contour plots of the joint analysis here. As elaborated in the main text, we can see that the slopes of the individual RTFs are correlated to a certain extent in the joint analysis, while the intrinsic scatters of the same have negligible correlation. Also, one can notice that the intrinsic scatter of the bin $\sigma_{n}(0.6\, \rm{Ropt})$ almost shows lower intrinsic scatter, validating that this radial bin is well constrained. Similarly, also the slope $a_{n}(0.6\, \rm{Ropt})$ can be seen to have negligible correlation with the rest of the RTFs slopes. 

\begin{figure}
    \centering
    \includegraphics[scale = 0.65]{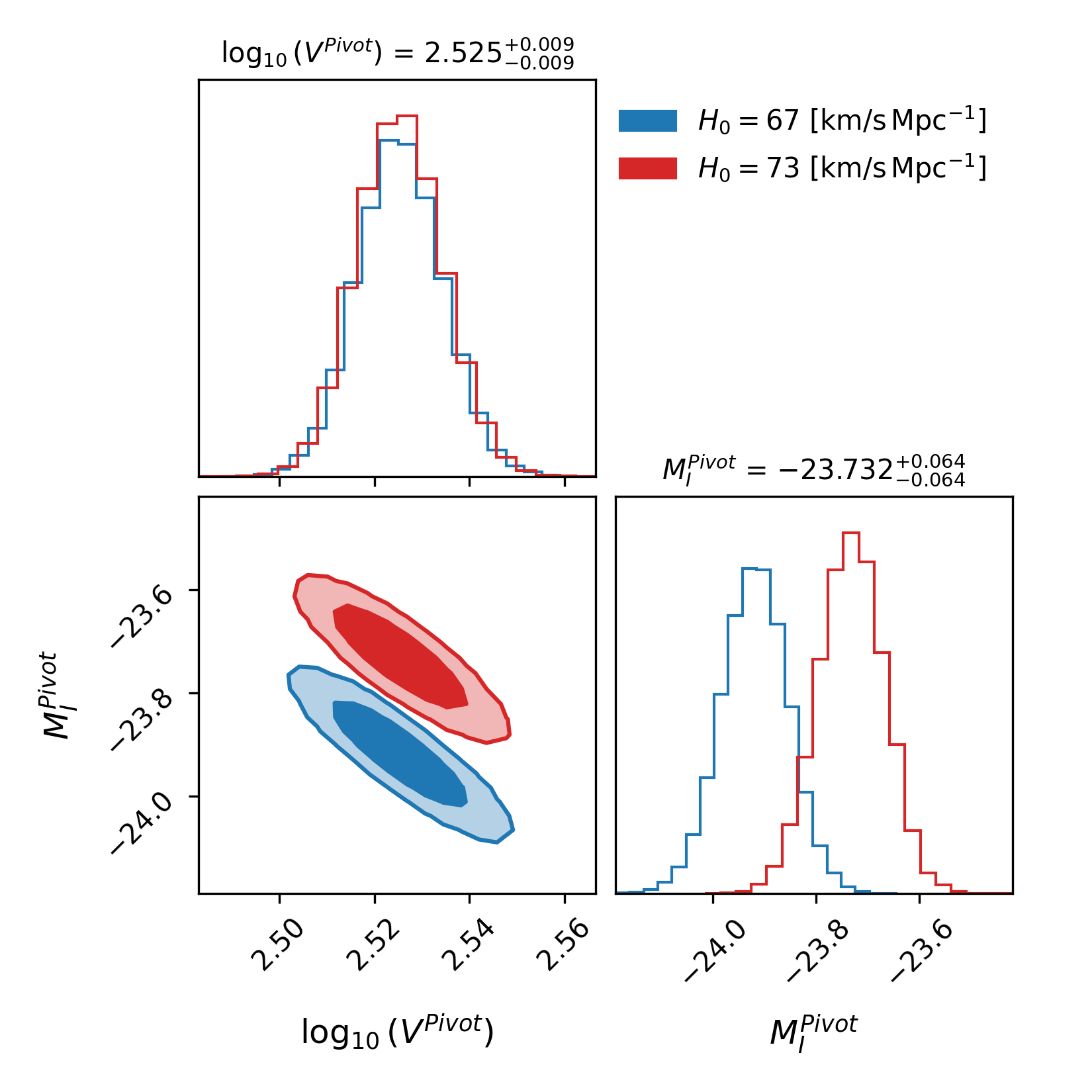}
    \caption{We show the $68\%$ and $95\%$ C.L. posteriors for the parameters $\logV$ and $\MI$, when the likelihood analysis is performed assuming $H_0 = 67 \, \ksM$ (blue) and $H_0 = 73\, \ksM$ (red). The $1\sigma$ parameter constraints for the case of $H_0 = 73 \, \ksM$ are shown on top of the 1D posteriors.  }
    \label{fig:H0-comp}
\end{figure}

\begin{figure}
    \centering
    \includegraphics[scale = 0.58]{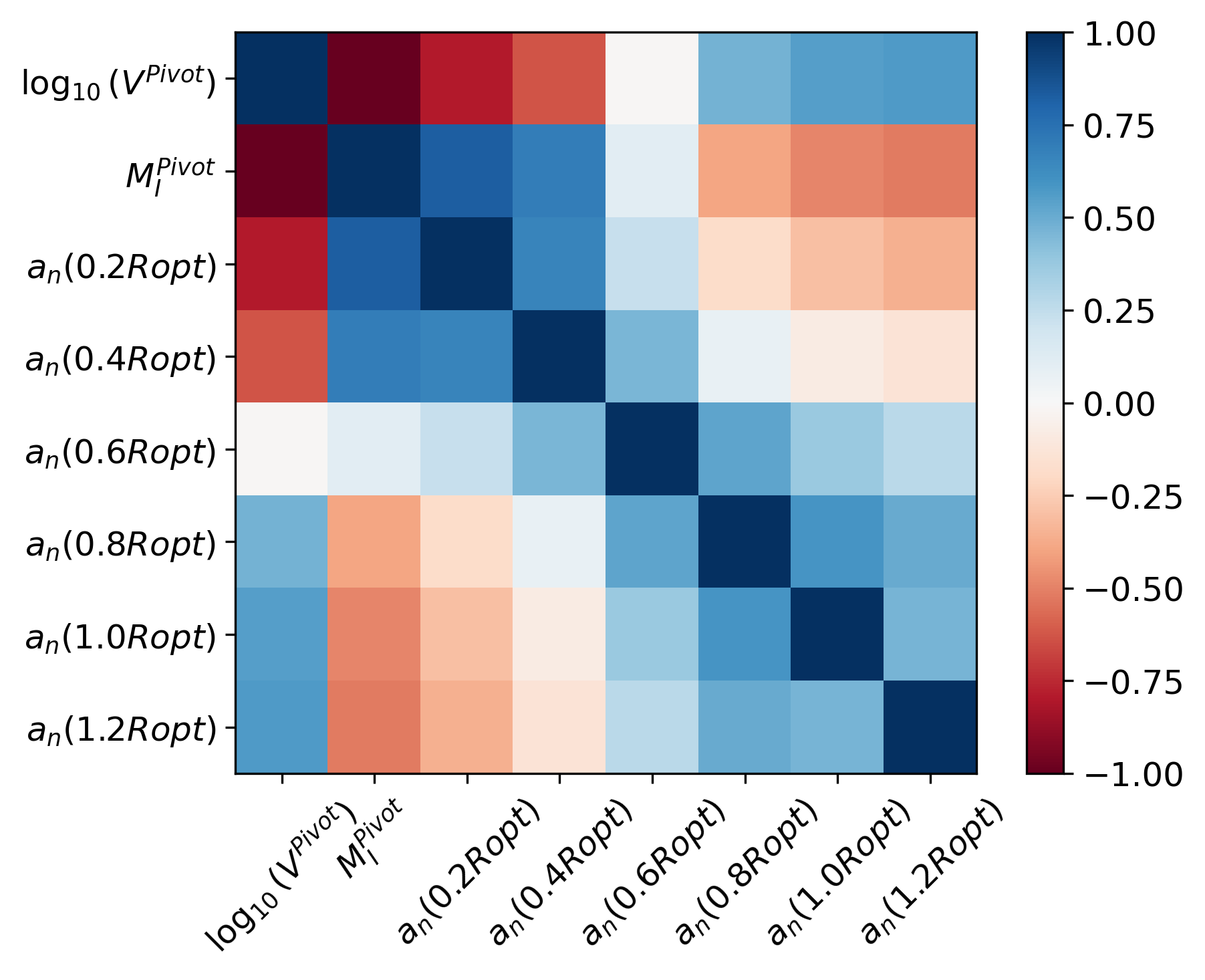}
    \caption{Correlation between the posteriors of the estimated slopes for the individual RTF relations and the pivot point.}
    \label{fig:Corr}
\end{figure}

\begin{figure*}
    \centering
    \includegraphics[scale = 0.24]{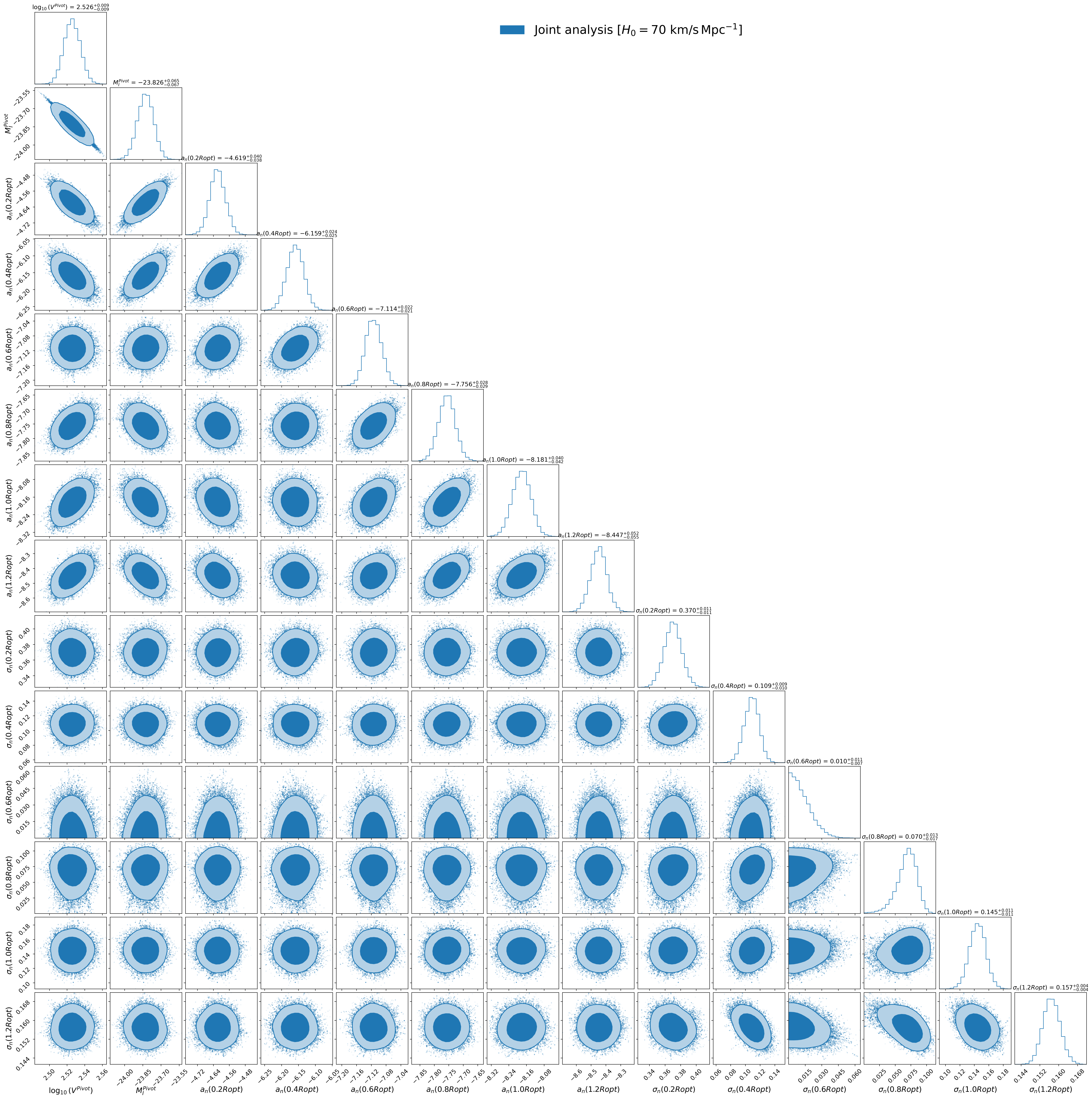}
    \caption{Contour plots showing the $68\%$ and $95\%$ C.L. for all the parameters in the joint analysis of the RTFs. We assume the $H_0 = 70\, \ksM$. }
    \label{fig:corner_full}
\end{figure*}

\bsp	
\label{lastpage}
\end{document}